\documentclass[5p]{elsarticle}

\usepackage{graphicx}
\usepackage{amsmath}
\usepackage{amssymb}
\usepackage{color}
\usepackage{ifpdf}
\usepackage{float}
\usepackage[utf8]{inputenc}
\usepackage{multirow}
\usepackage{rotating}
\usepackage{subfigure}

\usepackage{moresize}
\usepackage{url}
\usepackage{booktabs}
\usepackage{listings}   
\usepackage{paralist}    
\usepackage{wrapfig}    
\usepackage{multirow}
\usepackage{ifpdf}
\usepackage{xspace}
\usepackage{keyval}  
\usepackage{color}

\definecolor{listinggray}{gray}{0.95}
\definecolor{darkgray}{gray}{0.7}
\definecolor{commentgreen}{rgb}{0, 0.4, 0}
\definecolor{darkblue}{rgb}{0, 0, 0.4}
\definecolor{middleblue}{rgb}{0, 0, 0.7}
\definecolor{darkred}{rgb}{0.4, 0, 0}
\definecolor{brown}{rgb}{0.5, 0.5, 0}

\usepackage[normalem]{ulem}
\makeatletter
\def\cyanuwave{\bgroup \markoverwith{\lower3.5\p@\hbox{\sixly \textcolor{cyan}{\char58}}}\ULon}
\def\reduwave{\bgroup \markoverwith{\lower3.5\p@\hbox{\sixly \textcolor{red}{\char58}}}\ULon}
\def\blueuwave{\bgroup \markoverwith{\lower3.5\p@\hbox{\sixly \textcolor{blue}{\char58}}}\ULon}
\font\sixly=lasy6 
\makeatother

\newif\ifdraft
\ifdraft
\usepackage{xcolor}
\definecolor{ocolor}{rgb}{1,0,0.4}
\newcommand{\onote}[1]{ {\textcolor{ocolor} { (***Ole: #1) }}}
\newcommand{\terminology}[1]{ {\textcolor{red} {(Terminology used: \textbf{#1}) }}}

\newcommand{\jhanote}[1]{ {\textcolor{red} { ***shantenu: #1 }}}
\newcommand{\alnote}[1]{ {\textcolor{blue} { ***andreL: #1 }}}
\newcommand{\amnote}[1]{ {\textcolor{blue} { ***andreM: #1 }}}
\newcommand{\smnote}[1]{ {\textcolor{brown} { ***sharath: #1 }}}
\newcommand{\pmnote}[1]{ {\textcolor{brown} { ***Pradeep: #1 }}}
\newcommand{\msnote}[1]{ {\textcolor{cyan} { ***mark: #1 }}}
\newcommand{\mrnote}[1]{ {\textcolor{purple} { ***melissa: #1 }}}
\definecolor{orange}{rgb}{1,.5,0}
\newcommand{\aznote}[1]{ {\textcolor{orange} { ***ashley: #1 }}}
\definecolor{dandelion}{cmyk}{0,0.29,0.84,0}
\newcommand{\mtnote}[1]{ {\textcolor{dandelion} { ***matteo: #1 }}}
\newcommand{\note}[1]{ {\textcolor{magenta} { ***Note: #1 }}}
\else
\newcommand{\onote}[1]{}
\newcommand{\terminology}[1]{}

\newcommand{\alnote}[1]{}
\newcommand{\amnote}[1]{}
\newcommand{\athotanote}[1]{}
\newcommand{\smnote}[1]{}
\newcommand{\pmnote}[1]{}
\newcommand{\jhanote}[1]{}
\newcommand{\msnote}[1]{}
\newcommand{\mrnote}[1]{}
\newcommand{\aznote}[1]{}
\newcommand{\mtnote}[1]{}
\newcommand{\note}[1]{}
\fi

\newcommand{\pilot}{Pilot\xspace}
\newcommand{\pilots}{Pilots\xspace}
\newcommand{\pilotjob}{Pilot-Job\xspace}
\newcommand{\pilotjobs}{Pilot-Jobs\xspace}
\newcommand{\pilotcompute}{Pilot-Compute\xspace}
\newcommand{\pilotcomputedescription}{Pilot-Compute Description\xspace}
\newcommand{\pilotdescription}{Pilot-Description\xspace}
\newcommand{\pilotcomputes}{Pilot-Computes\xspace}
\newcommand{\pilotdata}{Pilot-Data\xspace}

\newcommand{\pilotdataservice}{Pilot-Data Service\xspace}
\newcommand{\pilotcomputeservice}{Pilot-Compute Service\xspace}
\newcommand{\computedataservice}{Compute-Data Service\xspace}
\newcommand{\computeunitdescription}{Compute-Unit Description\xspace}

\newcommand{\pd}{PD\xspace}

\newcommand{\computeunit}{Compute-Unit\xspace}
\newcommand{\computeunits}{Compute-Units\xspace}
\newcommand{\dataunit}{Data-Unit\xspace}
\newcommand{\dataunits}{Data-Units\xspace}
\newcommand{\du}{DU\xspace}
\newcommand{\dus}{DUs\xspace}
\newcommand{\dud}{DUD\xspace}
\newcommand{\cu}{CU\xspace}
\newcommand{\cus}{CUs\xspace}

\newcommand{\upp}{\vspace*{-0.5em}}

\newcommand{\irods}{iRODS\xspace}

\lstdefinestyle{myListing}{
  frame=single,   
  backgroundcolor=\color{listinggray},  
  language=C,       
  basicstyle=\ttfamily \footnotesize,
  breakautoindent=true,
  breaklines=true
  tabsize=2,
  captionpos=b,  
  aboveskip=0em,
  belowskip=-2em,
}      

\lstdefinestyle{myPythonListing}{
  frame=single,   
  backgroundcolor=\color{listinggray},  
  language=Python,       
  basicstyle=\ttfamily \scriptsize,
  breakautoindent=true,
  breaklines=true
  tabsize=2,
  captionpos=b,  
}

\ifpdf
\DeclareGraphicsExtensions{.pdf, .jpg, .tif}
\else
\DeclareGraphicsExtensions{.ps,  .eps, .jpg}
\fi

\usepackage{listings}
\usepackage{paralist}
\lstnewenvironment{code}[1][]%
{
\noindent
\minipage{1.0 \linewidth} 
\vspace{0.5\baselineskip}
\lstset{
    language=Python,
    frame=single,
    captionpos=b,
    stringstyle=\ttfamily,
    basicstyle=\scriptsize\ttfamily,
    showstringspaces=false,#1}
}
{\endminipage}

\defaultleftmargin{1em}{}{}{}
\begin{document}
\begin{frontmatter}
\title{Pilot-Data: An Abstraction for Distributed Data}

\author{Andre Luckow$^{1}$, Mark Santcroos$^{1,2}$, 
  Ashley Zebrowski$^{1}$, Shantenu Jha$^{1*}$\\[0.5em] 
{\emph{$^{(1)}$ \footnotesize{RADICAL, Rutgers University, Piscataway, NJ 08854, USA}}}\\
{\emph{$^{(2)}$ \footnotesize{Bioinformatics Laboratory, AMC, University of Amsterdam, NL}}}\\
 \emph{$^{(*)}$ \footnotesize{Contact Author: \texttt{shantenu.jha@rutgers.edu}}}}

\date{}

\begin{abstract} 
  Scientific problems that depend on processing large amounts of data
  require overcoming challenges in multiple areas: managing
  large-scale data distribution, controlling co-place\-ment and
  scheduling of data with compute resources, and storing,
  transferring, and managing large volumes of data.  Although there
  exist multiple approaches to addressing each of these challenges and the 
  complexity of distributed environments, an
  integrative approach is missing; furthermore, extending existing
  functionality or enabling interoperable capabilities remains
  difficult at best.  We propose the concept of {\it Pilot-Data} to
  address the fundamental challenges of co-placement and scheduling of
  data and compute in heterogeneous and distributed environments with
  interoperability and extensibility as first-order concerns.
  \pilotdata is an extension of the \pilotjob abstraction for
  supporting the management of data in conjunction with compute tasks.
  \pilotdata separates logical data units from physical storage,
  thereby providing the basis for efficient compute/data placement and
  scheduling.  In this paper, we discuss the design and implementation
  of the \pilotdata prototype, demonstrate its use by data-intensive
  applications on multiple production distributed cyberinfrastructure
  and illustrate the advantages arising from flexible execution modes
  enabled by \pilotdata. Our experiments utilize an implementation of
  Pilot-Data in conjunction with a scalable \pilotjob (BigJob) to
  establish the application performance that can be enabled by the use
  of \pilotdata.  We demonstrate how the concept of \pilotdata also
  provides the basis upon which to build tools and support
  capabilities like affinity which in turn can be used for advanced
  data-compute co-placement and scheduling. 
\end{abstract}
\end{frontmatter}

\section{Introduction} 

Data generated by scientific applications, instruments and sensors is
experiencing an exponential growth in volume, complexity and scale of
distribution, and has become a critical factor in many science
disciplines~\cite{hey2009}.  The ability to analyze prodigious volumes
of data requires flexible and novel ways to manage distributed data
and computations.  Furthermore, analytical insight is increasingly
dependent on integrating different and distributed data sources,
computational methods and computing resources.

Scientific data is often inherently distributed either because it is generated
by geographically dispersed data sources (e.\,g. instruments and/or
simulations) or it is needs to be distributed to facilitate further processing
of the data (as done e.\,g.\ in the LHC grid~\cite{lhcb}). Working with
distributed data involves many challenges beyond its storage and management. A
specific challenge is that of data and compute (processing) and the difficulty 
in being able to effectively and reliably manage
co-placement and scheduling is in part due to the challenges inherent in
coordination in distributed environments; it is compounded by an increasingly
rich, but complex heterogeneous data-cyberinfrastructure, characterized by
diverse storage, data management systems and multiple transfer
protocols/mechanisms.

Furthermore, tools and data-cyberinfrastructure have not been able to
address the need to integrate distributed compute and data resources
and capabilities~\cite{nsf_aci,Gray:2005:SDM:1107499.1107503}.  Many
solutions focus on either data or compute aspects, leaving it to the
application to integrate compute and data; in addition, most currently
available scientific applications still operate in legacy modes, in
that they often require manual data management (e.\,g.\ the stage-in
and out of files) and customized or application-specific scheduling.

Although these challenges have existed for a while, they are having
progressively greater impact on the performance and scalability of
scientific applications.  For example, Climate Modeling as performed
by the Earth System Grid Federation~\cite{6404471} is inherently a
distributed data problem.  The overall data generated and stored is
2-10 PB, of which the most frequently used data is 1-2 PB in size. The
data is generated by a distributed set of climate centers, and it is
stored in a distributed set of federated archives.  It is used by a
distributed set of users, who either run data analyses on a climate
center with which they are associated, or they gather data from the
ESGF to a local system for their analyses. Furthermore, data which is
generated over time causes real-time changes --- spatial and temporal; the 
scheduling of data analysis jobs needs to be
responsive to these spatio-temporal data changes.

To alleviate barriers to scalability and dynamic execution
modes, and impediments from an increasingly diverse and heterogeneous
infrastructure, some of the questions that must be addressed include:
(i) What are the right abstractions for coupling compute and data that
hold for a range of application types and infrastructures?  (ii) How
can these utilize existing and well-known abstractions and not require
whole-scale refactoring of applications and tools?  (iii) How can the
inherent heterogeneity and complexity of distributed
cyberinfrastructure be managed?  In addition to addressing the
challenge of providing interoperable, uniform access to heterogeneous
distributed cyberinfrastructure, how can these abstractions also be
used to provide the ability to reason about ``what'' and ``when'' to distribute 
as well as ``how''? Can these abstractions also enable effective
and novel execution modes for data-intensive applications? Whereas no single 
abstraction (or paper) can answer all of these
questions, in this paper we introduce \emph{\pilotdata (PD)} as a novel 
abstraction for data-intensive applications that addresses the first three 
questions, and equally importantly, outlines a research path to understanding 
other questions.

\jhanote{Introduction should provide two views of the concept.
  Pilot-Store: when we talk of infrastructure Pilot-Data: when we talk
  of application (dynamic) data possibly use notion of early and late
  binding? use notion of equivalence to task?}\alnote{without explicitly 
  mentioning pilotstore (would like to avoid another Pilot* this comes in my 
  opinion across well.)}

\pilotdata (PD) is an extension of the \pilotjob abstraction and supports the
management of data in conjunction with compute tasks. It provides flexible
placement and scheduling capabilities for data by separating the allocation of
physical storage and application-level data. \pilotdata is based on the
abstraction provided by \pilotjobs, which has a demonstrable record of
effective distributed resource utilization and supporting a broad range of
application types~\cite{pstar12},~\cite{saga_bigjob_condor_cloud}. We explore
how \pilotjobs and \pilotdata can be used to efficiently manage distributed
data and compute an a dynamic set of heterogeneous resources. For this purpose,
\pilotdata provides a simple and useful notion of distributed logical location
that from an application's perspective is invariant over the lifetime; thus it
supports both a decoupling in time and space (i.\,e., allowing late-binding)
between actual physical infrastructure and the application usage of that
infrastructure. The suggestion that \pilotdata is a conceptual abstraction for
distributed data is predicated upon the fact, that like any valid abstraction,
it must provide a range of applications with a unifying programming model and
usage mode. \pilotdata must thus retain the flexibility to be used with
different CI whilst not constrained to different specific modes of execution or
usage. As we will discuss, \pilotdata provides a general approach to
data-compute coupling, in that it is not constrained to a specific scheduling
algorithm or infrastructure. \pilotdata defines a minimal interface for 
resource acquisition and usage providing applications and frameworks full 
control and thus, a high flexibility, on how these resources are used. We will 
utilize \pilotdata to examine the general challenges and issues in the specific 
context of BWA-- a well known Next-Generation Sequencing (NGS) analysis 
application~\cite{Li:2010:FAL:1741823.1741825}. 

Our focus is on addressing the compute-data management and scheduling problem
in the context of {\it production} DCI and not research infrastructures.
Whether it be OSG/EGI (infrastructure with O(1000) sites) or XSEDE/PRACE
(infrastructure with O(10) sites) or clouds, the ability to reason about data
placement strategies (data replication and/or partitioning) and resource
allocation strategies (compute-to-data vs. data-to-compute), when to
offload/distribute is required. The realization and solution to these
high-level questions however vary significantly between infrastructures. This
points to the role of conceptual abstractions which enable reasoning without
having to worry about implementation details for a given capability. We
acknowledge that there exist multiple other challenges viz., data security,
data access rights and policy, and data semantics and consistency. These are
all important determinants of the ultimate usability and usage modes but we
will not consider them to be in scope of the work of this paper. Our decision
is in part explained by the fact that our work is ultimately aimed towards the
development of abstractions and middleware for production distributed
cyberinfrastructure (DCI) such as EGI~\cite{egi}, PRACE~\cite{prace},
XSEDE~\cite{xsede}, and OSG~\cite{1742-6596-78-1-012057}, which will be
agnostic to specific security and data-sharing policies.

This paper is structured as follows: in \S2, we provide the reader with a
better appreciation for the scope and context of our work with production
distributed infrastructures in mind. We discuss related work in \S3. \S4
presents a detailed overview of \pilotdata\ -- the concept, its relation to
\pilotjob and its implementation in BigJob. \S4 also introduces the Pilot-API
as means of providing a common interface to \pilotjobs and \pilotdata and
exposing the joint capabilities to support data-compute placement. In \S5 we
present the design and implementation of a \pilot-based workload management
service specifically designed for data-intensive applications. We design and
conduct a series of experiments in \S6 in order to establish and evaluate
\pilotdata as an abstraction for distributed data. We conclude with a
discussion of the main lessons learned as well as relevant and future issues.

\section{Infrastructure for Distributed Data}

The landscape of solutions that have been devised over the years to
address the challenges and requirements of distributed data is vast.
In this section we provide a brief discussion of relevant cyberinfrastructure
for supporting data-intensive applications. Further, we briefly survey data 
management in production DCI, such as XSEDE, OSG and EGI.

\subsection{Software Infrastructure for Data Management}

Traditionally HPC systems provided separated storage and compute systems, which
lead to various inefficiencies in particular with the increasing scale mainly
due to the fact that data always needs to get moved out of the storage system
in order to facilitate processing. Commonly, parallel file systems, e.\,g.\
Lustre~\cite{lustre} and GPFS~\cite{Schmuck:2002:GSF:1083323.1083349}, have 
been used to manage data in conjunction with parallel applications. While these 
system enable applications to access data via a standard POSIX interface, a 
drawback is that they do not expose data locality via this interfaces, i.\,e.\ 
commonly an application does not have control about file system internal data 
movements and caching.

With the emergence of distributed computing, different remote interfaces to
storage resources and data transfer have been developed, such as the Storage
Resource Manager (SRM)~\cite{srm-ogf}, GridFTP~\cite{ogf-gfd-20} or Globus
Online~\cite{10.1109/MIC.2011.64}. SRM is a type of storage service that
provides dynamic file management capabilities for shared storage resources via
a standardized interface. SRM is primarily designed as an access layer with a
logical namespace on top of different site-specific storage services. SRM aims
to hide the complexity of different low-level storage services, but does not
allow applications to control and reason about geographically distributed data.
Various implementations of SRM -- each optimized for a particular use case -
exist: dCache~\cite{conf/europar/FuhrmannG06}, Castor~\cite{castor},
StoRM~\cite{StoRM-chep06} and DPM~\cite{dpm} to name a few. While Castor and
dCache e.\,g.\ support the management of tape and disk storage hierarchies, DPM
and StoRM are more lightweight and focus on disk-based storage. SRM is heavily
utilized in HTC environments to accommodate storage and access to larger
volumes of data.

Several distributed data management systems have been built on top of these
low-level storage systems to facilitate the management of geographically
dispersed storage resources. The Global Federated Filesystem (GFFS)~\cite{gffs}
for example provides a global namespace on top of a heterogeneous set of
storage resources. Storage systems can be accessed via different mechanisms,
e.\,g.\ the virtual filesystem layer in Linux or a transfer protocol, such as
GridFTP.

\irods is a comprehensive distributed data management solution designed to
operate across geographically distributed, federated storage resources. \irods
~\cite{Rajasekar:2010:IPI:1855046} combine storage services with services for
metadata, replica, transfer management and scheduling. Central to \irods are
the so called micro-services, i.\,e.\ the user defined control logic.
Micro-services are automatically triggered and handle pre-defined tasks,
e.\,g.\ the replication of a data set to a set of resources. Also, different
services covering singular aspects such as replica management (e.\,g.\ the
Replica Location Service (RLS)~\cite{Chervenak:2004:PSR:1032647.1033304} or the
LCG File Catalogue (LFC)~\cite{lfc-1520941}) exist.

A main limitation of current infrastructures is the fact that they 
treat data and compute differently and require the user to (often painfully) 
manage data and compute resources separately. 
A reason for the limited number of higher-level services for data management
and integrated compute/data capabilities is the complexity and variety of
distributed applications make it difficult to foresee particular data access
pattern. Thus, file placement is mostly handled by the application and at best
supported by application-level services. Also, the available systems do not 
provide defined quality-of-service and applications are typically unaware of 
throughput and latencies to expect. Both limitations emphasizes the importance 
of higher-level application abstractions for distributed data/compute 
placements that enable applications to trade-off different aspects at runtime.

However, it also must be noted that some systems emerged that attempt to blur
the lines between compute and data. Hadoop~\cite{hadoop} for example aims to
address this issue by providing an integrated system for compute and data.
Hadoop is optimized for data-intensive, write-once/read-many and sequential
read workloads at the cost of Posix compliance. Also, it tightly couples
compute and data, i.\,e. the compute framework MapReduce is directly linked to
the underlying distributed filesystem Hadoop Filesystem (HDFS). A main 
limitation of Hadoop is the fact, that it is constrained to localized clusters 
and does not support distributed data very 
well~\cite{Cardosa:2011:EME:1996092.1996100}.

\subsection{Production Cyberinfrastructure}

\jhanote{Although there was discussion of individual tools and services, that
does not amount to an understanding of what integrated capabilities an
infrastructure such as Open Science Grid provides. We would like to possibly
use this opportunity to understand what is the status of data storage
capabilities in “distributed” infrastructure (including clouds, where there
might be dynamic provisioning of storage). A logical consequence of dynamic
data storage capabilities is the need for dynamic data placement capabilities.
We should try to survey both aspects".}\alnote{refined}

Data management has become an increasingly important task on production
infrastructures.  In this section we explore the status of data 
capabilities in ``distributed'' HPC and HTC infrastructure. HPC infrastructure, 
such as XSEDE~\cite{xsede}, are
primarily concerned with compute-intensive tasks and thus, lack some
distributed data/compute management services that are provided in HTC
environments, such OSG~\cite{Altunay:2011:SDP:1997543.1997573} and
EGI~\cite{egi}.

\begin{table}[t]
	\scriptsize
	\centering
\begin{tabular}{|p{1cm}|p{1.4cm}|p{1.4cm}|p{1.4cm}|p{1.4cm}|}
		\hline
  	  			       &\textbf{XSEDE}        &\textbf{OSG}               
					   &\textbf{EGI}    &\textbf{Atlas/OSG} \\
		\hline
		Storage        &Local, Parallel Filesystems  &Local, SRM, \irods        
		  			   &Local, SRM      &Local, SRM\\
		\hline
		Data Access    &SSH, GridFTP, Globus Online          &SSH, SRM, \irods           &SSH, SRM        &SSH, SRM, XROOTD\\
		\hline
		Manage\-ment     &Manual              &Manual, \irods, BDII             
		&Manual, BDII
																							&XROOTD, PD2P\\
		\hline
\end{tabular}
\caption{Data-Cyberinfrastructure \label{table:data-cyberinfrastructure}}
\end{table}

Table~\ref{table:data-cyberinfrastructure} summarizes the data
cyberinfrastructures deployed by the different production DCIs. The landscape
of data cyberinfrastructure is very heterogeneous with most applications only
utilizing local capabilities, e.\,g.\ parallel filesystems. With the increasing
need for supporting distributed data and compute, infrastructures started to
deploy more sophisticated capabilities, e.\,g.\ XSEDE and OSG provide \irods
support for some resources. In particular, on high-throughput infrastructures
distributed data access is essential since parallel filesystems are commonly
not provided. Thus, a myriad of data management options emerged on HTC 
infrastructures. Historically, EGI and OSG introduced SRM as unified access 
layer for storage resource pools co-located to their compute resources. In 
addition, OSG  provides with \irods a higher-level data
management services, which supports simple means of managing data in a highly
distributed environment of compute resources. For example, data replication can
be used to replicate a datasets to a group of resources to facilitate
distributed computation at a later stage. However, the application is still
required to manually manage the mapping between these compute and storage
resources.

Further infrastructures provide service for managing meta-data and information
for data/compute resources. EGI and OSG e.\,g.\ offer services for replica
management and for resource information (BDII~\cite{bdii}. However, the
application is then required to combine these building blocks and to manually
construct a resource topology. A reason for this is that the complexity and
variety of distributed applications make it difficult to foresee particular
data access pattern. Thus, file placement is mostly handled by the application
and at best supported by application-level services. A notable exception a
domain specific infrastructures, e.\,g.\ the LHC Grid. Nevertheless, this
emphasizes the importance of generic, higher-level application abstractions for
distributed data/compute placements that can serve broader communities.

In addition to the described services, several domain-specific and higher-level
approaches for distributed compute/data management emerged. For example, in
context of the Atlas collaboration~\cite{1742-6596-119-6-062036}, various
tools for managing distributed data have been developed
on top of OSG. Xrootd~\cite{xrootd} e.\,g.\ is a distributed storage system
that is capable of managing data across geographically dispersed resources
using a hierarchical management structure. Based on this lower-level
infrastructure various tools for managing computed data exists: the ROOT and
PROOF systems e.\,g.\ enable the analysis of data stored in Xrootd. Further, as
part of this infrastructure the \pilots (e.\,g.\
PanDA~\cite{1742-6596-119-6-062036}) are used to manage compute in conjunction
with datasets residing on SRM or Xrootd (see section~\ref{sec:related}).

In the cloud space a separate ecosystem of storage services emerged. 
A novel type of storage introduced by cloud environments are object stores, a
form of highly distributed storage that can potentially be distributed across
multiple data centers. Object stores are optimized primarily for ``write once,
read many'' workloads and can support massive volumes of data with their
scale-out architectures. For example, Amazon S3~\cite{amazons3} automatically
replicates data across multiple data centers within a region. These kind of
stores are not suitable for all workloads (e.\,g.\ traditional, transactional
workloads). On the other hand, typical Big Data workloads that (i) require the
storage of large volumes of data and (ii) are characterized by a large amount
of reads are particularly suitable for such stores. Access to such storage
systems is via a common -- often simplified -- namespace and API. For example,
cloud systems, such as the Azure Blob Storage, Amazon S3 and Google Storage,
provide only a namespace with a 1-level hierarchy. This means that applications
need to be adapted, in order to benefit from object storage. The most widely
used object stores are: Amazon S3~\cite{amazons3}, Azure
Storage~\cite{azure-blob-storage} and Google Cloud
Storage~\cite{google-storage}. In addition, both Eucalyptus and OpenStack
provide an object store: Eucalyptus Walrus~\cite{walrus} and OpenStack
Swift~\cite{openstack-swift}. A major limiting factor is the necessity to
ingest large volumes of data to the cloud storage over the WAN. Large volume
data transfers are associated with high costs and unpredictable and/or
unacceptable performance. Also, data typically has to be moved to the compute
resource (usually a VM) for processing.

While there are various useful services and building blocks for data-intensive
application available, they typically require the application to utilize
specialized access libraries and tools as well as to manually manage
compute/data co-placements by providing the right resource constraints to the
scheduler. Higher-level abstractions for compute and data and smart
data/compute placement services are missing capabilities of existing
infrastructures. While data placement strategies are extensively investigated
(see e.g.~\cite{Ma20131395}), currently most production DCI do not support
distributed data placements. Having integrated compute/data capabilities and
the ability to manage dynamic compute/storage resources is an essential
requirement for effectively supporting and scaling dynamic and distributed
applications on production infrastructures overcoming currently prevailing
inflexible execution model.

\section{Related Work}
\label{sec:related}

In this section, we explore related work with respect to (i) existing systems
and algorithms for managing distributed data and compute, (ii) abstractions and
programming models for data-intensive applications, (iii) data management in
the context of \pilotjobs.

\noindent
\emph{Distributed Data/Compute Management Systems and Algorithms:} Managing distributed data and 
compute has been an ongoing research topic. For
grid environments for example, several For example, the Stork~\cite{1281599}
data-aware batch scheduler provides advanced data and compute placement for
Condor and DAGMan. Stork supports multiple transfer protocols like, SRM,
(Grid)FTP, HTTP and SRB. Romosan et al.~\cite{Romosan05co-schedulingof} present
another data-compute co-scheduling approach on top of Condor and SRM. Both
approaches build on top of existing job scheduling and data-transfer and
storage solutions.
Further frameworks for other distributed environments have been proposed.
FRIEDA~\cite{10.1109/SC.Companion.2012.132} for example provides a data
management framework for cloud-environments.

Different research on when to (potentially dynamically) distribute and
replicate data has been conducted: for example, Foster~\cite{1029935} and
Bell~\cite{Bell01112003} investigate different data replication management
system and dynamic replication algorithms in the context of scientific data
grids. A limitation of the previous approaches is that the systems and
algorithms are usually constrained to system-level replication, making it
difficult for the user to control replication on application-level and employ
dynamic replication strategies. Glatard et\,al.~\cite{Ma20131395} attempt
to provide a classification of data placement and replications algorithms and
systems for distributed environments.

\noindent
\emph{Abstractions and Programming models:} Various abstractions for optimizing
access and management of distributed data have been proposed:
File\-cule~\cite{1652137} is an abstraction that groups a set of files that are
often used together, allowing an efficient management of data using bulk
operations. This includes the scheduling of data transfers and/or replications.
Similar file grouping mechanisms have been proposed by Amer
et\,al.~\cite{1022302}, Ganger et\,al.~\cite{Ganger97embeddedinodes} and
BitDew~\cite{Fedak:2008:BPE:1413370.1413416}. Further several higher-level,
less resource-oriented abstractions for enabling data analysis on large volumes
of data have been proposed. A well-known example is the MapReduce programming
model~\cite{mapreduce} for which various implementations 
exist~\cite{hadoop,Mantha:2012:PEF:2287016.2287020}. Another example is 
Data\-Cutter~\cite{Beynon:2001:DPV:543586.543590}, a
framework that enables exploration and querying of large datasets while
minimizing the necessary data movements. While various abstractions for 
data-intensive applications exist, these are typically bound to a specific 
infrastructure. For example, Hadoop -- the most-widely used MapReduce 
implementation -- intermingles resource management, programming abstraction in 
a monolithic solution sacrificing flexibility and extensibility with respect to 
other kinds of data-intensive workloads. \pilotjobs have been shown to being 
capable of providing a flexible and extensible resource management layer to 
different types of application workloads. Increasingly, data-intensive 
applications are being supported on top of a \pilotjob.

\noindent
\emph{Data Management and \pilotjobs:} \pilotjobs have been successful
abstractions in distributed computing as evidenced by a plethora of PJ
frameworks. With the increasing importance of data, \pilotjobs have been also
used to process and analyze large data. However, in most \pilotjob framework
the support for data movement and placement is insufficient~\cite{pstar12}.
Only a few of them provide integrated compute/data capabilities, and where they
exist, they are often non-extensible and bound to a particular infrastructure.
In general, one can distinguish two kinds of data management: (i) the ability
to stage-in/stage-out files from another compute node or a storage backend,
such as SRM and (ii) the provisioning of integrated data/compute management
mechanisms. An example for (i) is Condor-G/Glide-in, which provides a basic
mechanism for file staging and also supports access to SRM. Another example is
Swift~\cite{Wilde2011}, which provides a data management component called
Collective Data Management (CDM). DIANE provides in-band data transfer
functionality over its CORBA channel.

In the context of the LHC Grid several type (ii) \pilotjob frameworks that
support access to the vast amounts of experimental data created by the Large
Hadron Collider have been developed. DIRAC~\cite{Tsaregorodtsev:2010zz} is an 
example of such a system. It interfaces to SRM storage resources and enables the
application to stage-in/out data to this system.
AliEn~\cite{1742-6596-119-6-062012} also provides the ability to tightly
integrate storage and compute resources and is also able to manage file
replicas. While all data can be accessed from anywhere, the scheduler is aware
of data localities and attempts to schedule compute close to the data.
Similarly, PanDA~\cite{panda-overview} provides support for the retrieval of
data from the Xrootd storage infrastructure. The PanDA Dynamic Data Placement
component~\cite{maeno_pd2p:_2012} provides a demand-based replication system,
which can replicate popular datasets to underutilized resources for later
computations. However, this capability is provided on system-level and
constrained to official Atlas datasets, i.\,e.\ it cannot be applied to
user-level datasets. The data/compute management capabilities of
DIRAC, AliEn and PanDA are built on top of Condor-G/Glide-in. In addition to 
this strong coupling to the underlying infrastructure, these 
frameworks are tightly bound to their specific applications.

Another example for a type (ii) system is Falkon~\cite{1362680}, which provides
a data-aware scheduler on top of a pool of dynamically acquired compute and
data resources~\cite{Raicu:2008:ALD:1383519.1383521}. The so called data
diffusion mechanism automatically caches data on \pilot-level enabling the
efficient re-use of data. Falkon provides limited interoperability and is 
constrained to Globus-based grid environments.

As alluded before MapReduce is a popular abstraction for expressing
data-intensive, analytical applications. Hadoop is the most widely-used
implementation of MapReduce and provides a vertical stack consisting of a
distributed filesystems, a data-aware job scheduler, the MapReduce framework as
well as various other frameworks. With the increasing variety of Hadoop-based
applications and frameworks, the requirements with respect to resource 
management increased, e.\,g.\ it became a necessity to support batch, streaming 
and interactive data processing. YARN~\cite{yarn} is the new central resource 
manager of Hadoop 2.0 that was developed to address these needs. While YARN
solves some of these problems, it has some limitations: it provides e.\,g.\
only a very low-level abstraction for resource management; data locality needs
to be manually managed by the application by requesting resources at the
location of an file chunk. With the need for supporting even more heterogeneous
workloads, \pilot-like frameworks for Hadoop emerged, e.\,g.\
Llama~\cite{llama} and Tez~\cite{tez}. Another scheduler proposed for Hadoop is
Mesos~\cite{Hindman:2011:MPF:1972457.1972488}: in contrast to YARN, Mesos is a
two level scheduler, i.\,e.\ similar to \pilots resources are initial requested
and afterwards directly managed by the application.

\section{Pilot-Data: An Unified Abstraction for Compute and Data}

\emph{\pilotdata} was conceived as a unified abstraction to distributed
data management in conjunction with \pilotjobs -- a capability that has 
been neglected by many \pilotjob frameworks.  \emph{A \pilotjob is defined as
an abstraction that generalizes the reoccurring concept of utilizing a
placeholder job as a container for a set of compute tasks~\cite{pstar12}.}
From a practical point-of-view, data management and movement for most
\pilotjobs -- if it exists at all -- is at best ad hoc and not generic.
Consequently most \pilotjobs rely on application-level data management, i.\,e.\
data needs to be pre-staged or each task is responsible for pulling in the data.

\emph{\pilotdata (PD) is an extension of the Pilot-Job abstraction for
  supporting the management of data in conjunction with compute
  tasks. \pd separates logical compute and data from physical resource
  enabling efficient compute/data placements using various strategies
  (e.\,g.\ moving compute to data and vice versa, opportunistic
  replication, partitioning) independent from the underlying infrastructure.} 
Pilot-Data provides a well-defined semantic for data movement, storage and 
access in conjunction with compute carried out through Pilot-Jobs.  Pilot-Data 
separates application-level data and compute tasks from infrastructure-level 
storage/compute enabling efficient compute/data placements. Like \pilotjobs, it 
allows for application-level control, and the logical decoupling of physical 
storage/data locations from the production and consumption of data. In summary, \pilotdata provides the following key capabilities:
\begin{compactenum}
	\item\emph{Dynamic Resource Management:} It supports the management and 
	access to (dynamic) storage resources 
	in conjunction with Pilot-Jobs. For this purpose, it provides a unified 
	access layer to different heterogeneous data cyberinfrastructure, such as: 
	SRM, iRODS, Globus Online, S3. \pilotdata facilitates
	and utilizes late binding of data and physical resources for optimal
	coupling and management. The framework is agnostic to the type of data 
	and can be used to managed arbitrary data in distributed, dynamic 
	environments.
	
	\item\emph{Distributed Namespace for Data:} \pilotdata provides a simple, 
	two-level distributed, global namespace spawning heterogeneous storage 
	resources that can be accessed from any resource.

	\item\emph{Higher-Level Abstraction for Compute/Data Coupling:} The 
	\pilotdata abstraction aims to support the 
	coupling of different application components and the management of the data 
	flow between the different application stages, e.\,g.\ the parts of a 
	distributed workflow.  The \dataunit abstraction provides the ability to 
	group files. The \dataunit URL serves as a 
	single level namespace independent of the actual physical location of the 
	\dataunit, which can be e.\,g.\ replicated across multiple geographically 
	distributed \pilotdata. Further, the 
	application is able to organize files into an application-level 
	hierarchical namespace within a \dataunit.

	\item \emph{Compute/Data Scheduling:} \pilotdata supports the co-scheduling 
	of compute and data into co-located \pilotcompute and \pilotdata using  a 
	data-aware workload management service. Also, applications can utilize 
	lower-level primitives (i.\,e.\ the \pilot-Abstraction) to manually 
	optimize data and compute placements.
\end{compactenum}
\pilotdata combines a unified abstraction for compute and data with an 
interoperable \pilot-based access layer to heterogeneous resources bridging 
often disperse compute and data cyberinfrastructures allowing applications to 
reason about data/compute universally.

\subsection{Design Objectives}

In this section we derive the design objectives for \pilotdata based on common
data and processing patterns. Commonly applications deal with two types of
data: static and dynamic data. The majority of data is static and resides in
archives to which it is written once, but never modified; most of this data
remains unanalyzed~\cite{Gantz:2012uq}. Commonly this type of data is shared
between multiple users and applications. \pilotdata can provide a unified
abstraction for accessing these datasets using compute resources allocated via
a \pilot. Dynamic data arises when some property of the input data or its
delivery changes, for example in terms of arrival rate, provenance, burstiness,
or source. There may be variability in the structure of the data, for example,
data schema, file formats, ontologies, etc~\cite{jha:2011fk}. A good example
are data feeds generated by scientific instruments, experiments, simulations or
dynamic workflows. With the dynamic nature of data the requirements with
respect to compute grow: dynamic data comes with dynamic and potentially
realtime processing requirements. In this case the lifecycle of the data is
tightly bound to the lifecycle of compute. Thus, an integrative approach of
compute/data management is essential. \pilotdata can be used to allocate
appropriate storage resources to meet the space and I/O requirements for
dynamic data and facilitates the effective processing of this data in a
\pilotcompute.

Typically, scientific applications involve multiple steps of data generation
and processing. Examples of application patterns are: ensembles, coupled
ensembles, more complex pipelines possible comprising of different kinds of raw
and derived data~\cite{Gray:2005:SDM:1107499.1107503}, MapReduce-based
applications and workflows. Often, input data is partitioned to facilitate the
data-parallel processing of data, e.\,g. by an ensemble of tasks. In this case
on can differentiate between (i) partitioned data, i.\,e.\ data that is divided
in a way so that each task consumes a unique part of the data, and (ii) shared
data that is required by all tasks. Dynamic data often arisen in multi-stage 
workflows where it is often difficult to predict the output of the previous 
stage.

\pilotjobs have been shown to be highly effective in supporting fine-grained
ensemble tasks and proofed particular useful for applications with dynamic
resource requirements. \pilotdata aims to address (i) the data management
challenge arising from data-intensive applications and (ii) the potentially
dynamic compute/data requirements associated with the needs of dynamic and
distributed compute/data trading of properties such as storage type (and
associated IOPS), network bandwidth, compute capacity and data locality.
Assuming a dynamic data feed from an scientific instrument as an example. If
sufficient storage and bandwidth to the storage is available, data can be
cached and then processed in a \pilotcompute. If the application
has realtime requirements, i.\,e.\ a latency requirements with respect to
availability of results relative to data volume and computational complexity,
resource management becomes even more challenging. Usually, this requires the
usage of readily available compute resources from a pool of \pilots.
In summary, \pilotdata is designed to address the following usage modes:
\begin{compactenum}
	\item Manage input and output for \pilot-based applications and provide 
	access to user-owned as well as community datasets available on many 
	infrastructures. \pilotdata supports applications in exploring data 
	parallelism, e.\,g. by allowing them to efficiently partition and/or 
	replicate data. This way it allows applications to optimize data movements, 
	e.\,g.\ to create a \pilot-local replica of dataset to facilitate the 
	faster processing of this data.
	\item Manage \emph{dynamic} data in different scenarios, e.\,g. within
	data-intensive, dynamic workflows or the intermediate data within 
	MapReduce. In this cases it is necessary to create short-term, transient 
	``storage space'' for intermediate data, which can be removed after the end 
	of the application run.
	\pilotdata enables the reasoning about application and resources, such as
	data/compute localities and placements on top of a dynamic compute/data
	overlay.\label{pilot-analogy}
	\item Support common data processing patterns, such as data-partitioning, 
	parallel processing and output gathering.
\end{compactenum}

\noindent
While there are many commonalities between compute and data, depending on the
type of data there are distinct differences: the lifecycle of static data
extremely differs from compute -- data commonly outlives computes and is often
consumed by different kinds of compute. Nevertheless, the management of such
data sets is a challenge: commonly, these datasets are partitioned, filtered
and replicated in user space using a myriad of scripts making it difficult
track transformations and results. The lifecycle of transient, dynamic data is
commonly strongly coupled to the lifecycle of the associated compute.
\pilotdata enables \pilot-based applications to acquire appropriate data
resources needed for processing their datasets. It further enables them to
manage file movements to and from these resources. By doing so, \pilotdata
allows \pilot-based applications to utilize data locality in conjunction with
their \pilotcomputes. Applications e.\,g.\ can create dynamic caches in
conjunction with their \pilotcomputes, which can be used for intermediate data
or for the fan-out additional compute tasks.

\subsection{BigJob: A Pilot-Compute and Data Implementation}

\begin{figure}[t]
	\centering
	\includegraphics[width=0.45\textwidth]{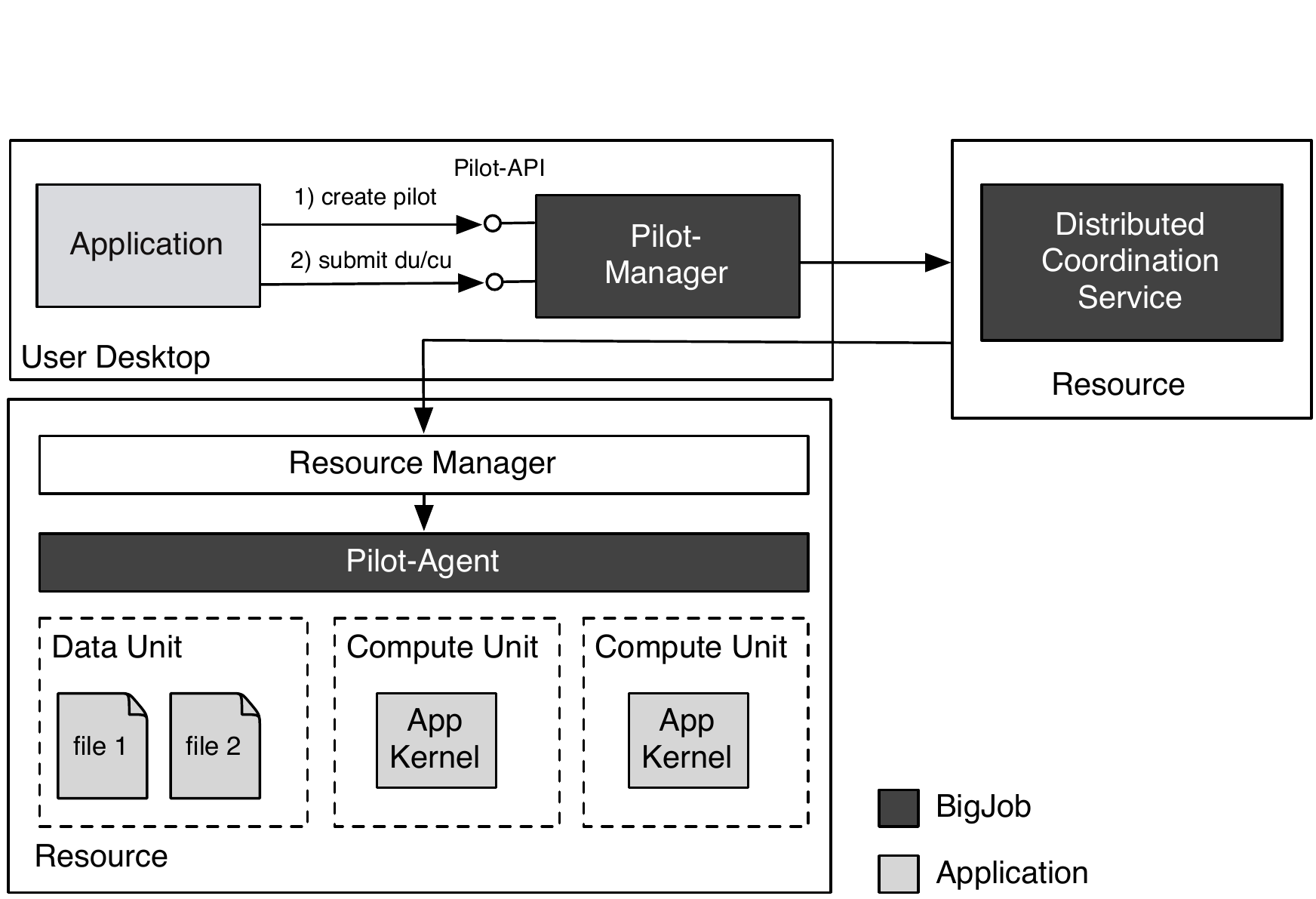}
	\caption{\textbf{BigJob High-Level Architecture:} The \pilot-Manager is 
	the central coordinator of the framework, which orchestrates a set of 
	\pilots. Each \pilot is represented by a decentral component referred to 
	as the \pilot-Agent, which manages the set of resources assigned to it.\upp\upp\upp}
	\label{fig:figures_bigjob-bigdata-architecture}
\end{figure}

{\it Architecture and Design: } Consistent with our aims of providing
complete \pilotjob capabilities, we implement \pilotdata as an extension of
BigJob (BJ) ~\cite{saga_bigjob_condor_cloud,Mantha:2012:PEF:2287016.2287020},
which is a SAGA-based \pilotjob implementation. BigJob provides an unified runtime environment for \pilotcomputes and \pilotdata
on heterogeneous infrastructures. The framework offers a higher-level,
interface -- the \pilot-API -- to heterogeneous and/or distributed data
and compute resources.
Figure~\ref{fig:figures_bigjob-bigdata-architecture} shows the high-level
architecture of BigJob. The \pilot-Manager is the central entity of the 
framework, which is responsible for managing the lifecycle of a set of \pilots 
(both \pilotcomputes and \pilotdata).  For this purpose BigJob relies on a set 
of resource adaptors (see adaptor pattern~\cite{gamma94}).

A resource adaptor encapsulates the different infra\-structure-specific
semantics of the backend system, e.\,g.\ in the case of \pilotcompute different 
resource management systems and in the case of \pilotdata different
storage types (e.\,g.\ file vs. object storage), access and transfer protocols.
Using this architecture, BigJob eliminates the need for application developers
to interact directly with different kinds of compute and storage resources,
such as the batch queue of HPC/HTC resources or the VM management system of 
cloud resources.
 
\begin{figure}[t]
    \upp
	\centering
        \includegraphics[width=0.45\textwidth]{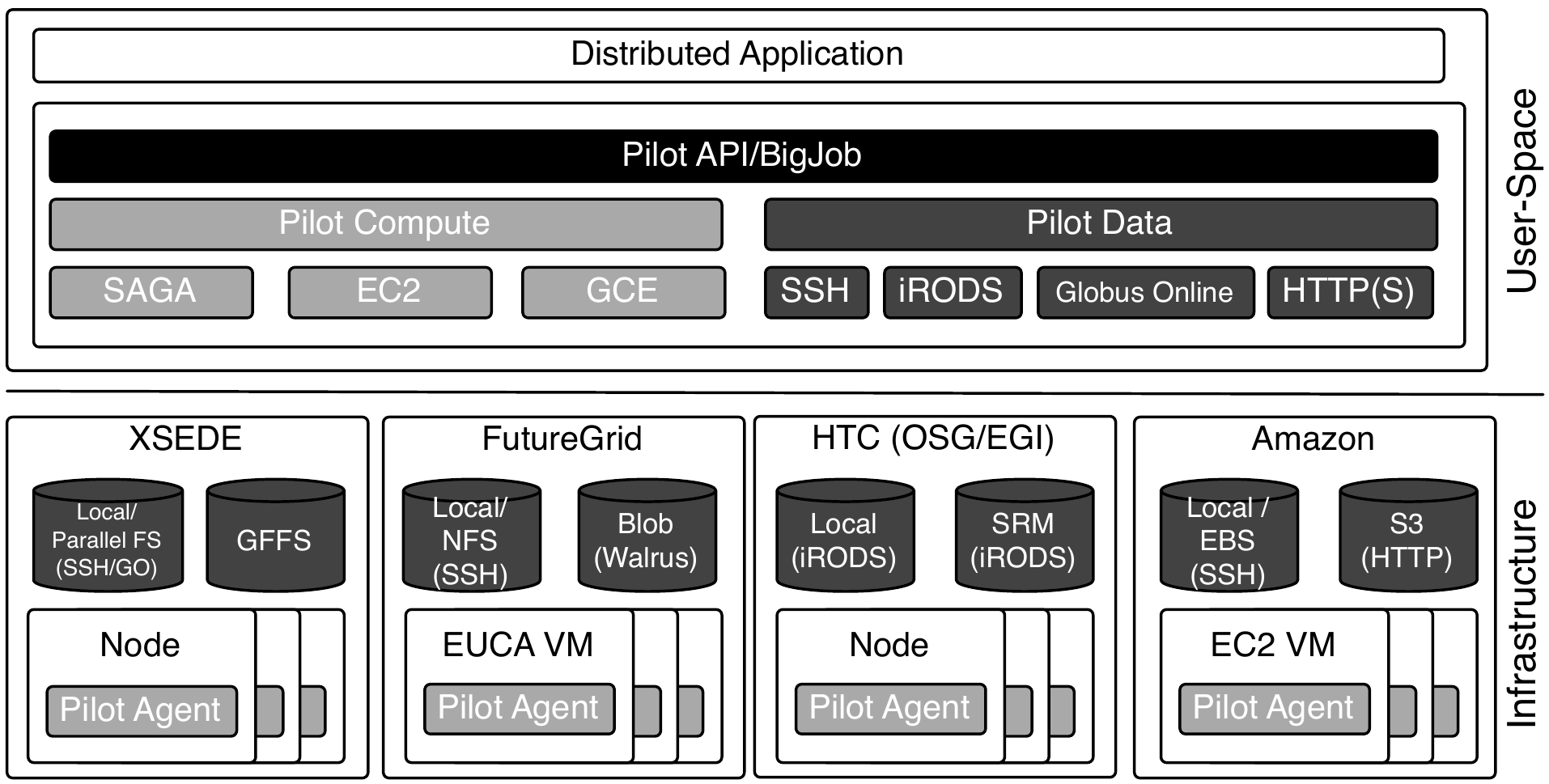}
        \caption{\textbf{BigJob Pilot Abstractions and Supported
            Resource Types:} BigJob provide a unified abstraction to a
          heterogeneous set of distributed compute and data
          resources. Resources are either accessed via
          SAGA~\cite{ogf-gfd-90,saga-python-pd} or via a custom
          adaptor.\upp}
	\label{fig:figures_cloud_pilot_job}
\end{figure}

As shown in Figure~\ref{fig:figures_cloud_pilot_job} BJ supports various types
of HPC/HTC resources via SAGA-Python~\cite{saga-python-pd} (e.\,g.\ Globus,
Torque or Condor resources). Further, adaptors for cloud resources (Amazon EC2
and Google Compute Engine) exist. A \pilotdata backend is defined by (i) the
storage resource and (ii) the access protocol to this storage. On XSEDE,
storage resources such as parallel filesystems (commonly Lustre or GPFS) can be
remotely accessed using different protocols and services, e.\,g.\ SSH, GSISSH,
GridFTP~\cite{ogf-gfd-20} and Globus Online~\cite{10.1109/MIC.2011.64}. Other
storages types, e.\,g.\ cloud object stores as S3 or \irods, tightly integrate
storage and access protocol and provide additional features such as data
replication. Each \pilotdata adaptor encapsulates a particular storage type and
access protocol.

{\it Runtime Interactions: } \pilots are described using a
JSON-based description (see \pilot-API in section~\ref{sec:pilot-api}), which 
is submitted to the \pilot-Manager. The
description contains various attributes that are used for expressing
the resource requirements of the \pilot. An important attribute is the
backend URL of the resource manager (for \pilotcomputes) or
the storage/transfer service (for \pilotdata). The URL scheme is used
to select an appropriate BigJob adaptor. Once an adaptor is
instantiated, it is bound to the respective \pilot object; all
resource specific aspects for this \pilot are then handled by this
adaptor. 

\begin{figure}[t]
	\centering
	\includegraphics[width=0.5\textwidth]{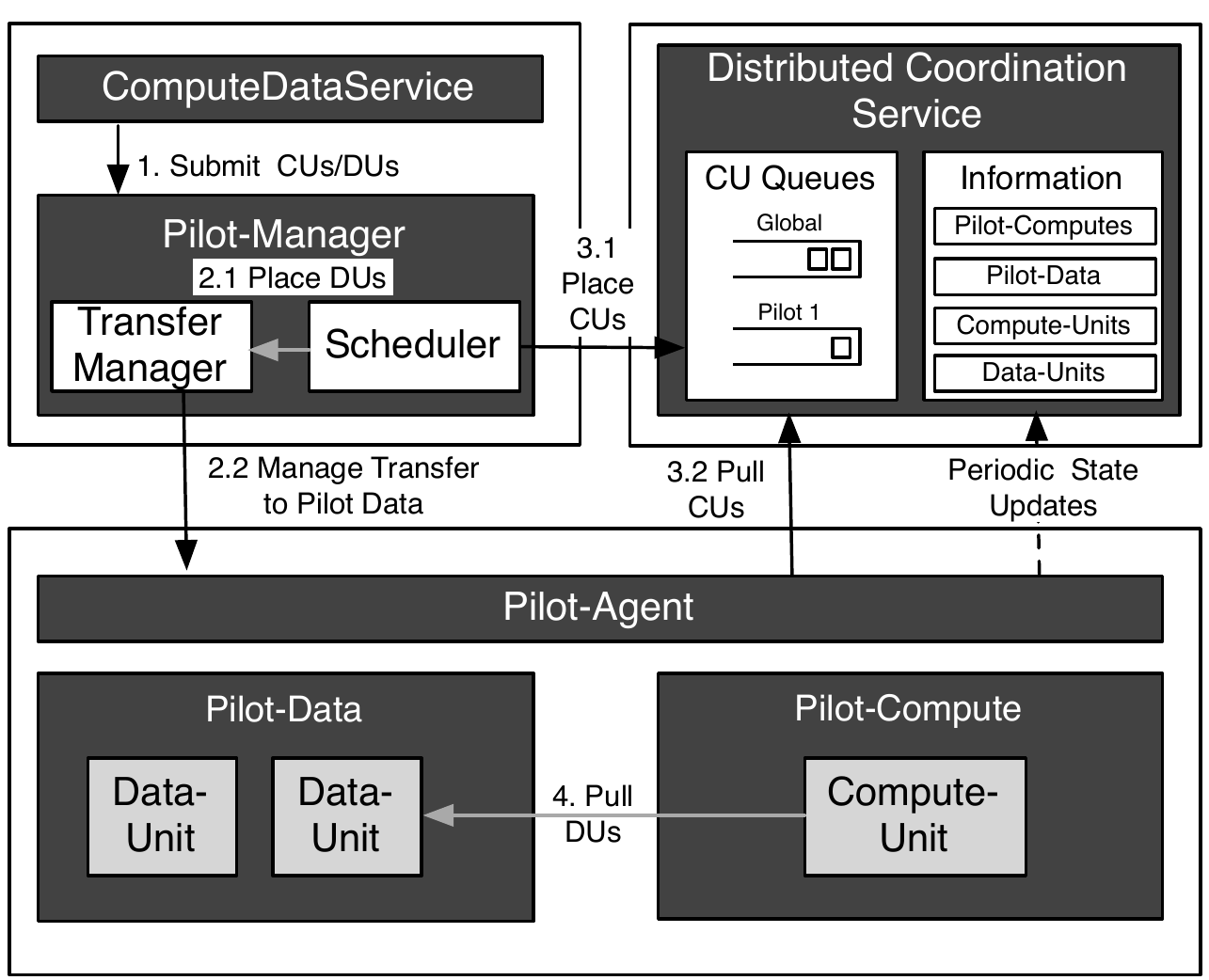}
	\caption{\textbf{BigJob Application Workload Management:} The figure
          illustrates the typical steps involved for placing and
          managing the application workload, i.\,e.\ the \dus and
          \cus.\alnote{sequence diagram to make figure 3 more readable? extract 
		  interactions to other figure.}}
	\label{fig:figures_computedataservice-scheduling}
\end{figure}

Figure~\ref{fig:figures_computedataservice-scheduling} shows the
typical interactions between the components of the BigJob/Pilot-Data
framework after the submission of the application workload (i.\,e.\
the \cus and \dus). The core of the framework is the
\pilot-Manager. The \pilot-Manager is able to manage multiple
\pilot-Agents. The application workload is submitted to the
\pilot-Manager via the \computedataservice interface of the \pilot-API
(see section~\ref{sec:pilot-api}). After submission, \dus and \cus are
put into a in-memory queue of the distributed coordination service,
which is continuously processed by the scheduler component. This
asynchronous interface ensures that the application can continue
without needing to wait for BigJob to finish the placement of a \cu or
\du.

{\it Distributed Coordination and Control Management:} The main task of the 
coordination \& communication service is to facilitate control flow and data 
exchange between distributed components of the framework, i.\,e.\ the 
\pilot-Manager and \pilot-Agent. BigJob uses a shared in-memory data store, 
Redis~\cite{redis}, for this purpose. 
Both manager and agent exchange various types of control data via a defined set
of Redis data structures and protocols: (i) The \pilot-Agent collects various
information about the local resource, which is is pushed to the Redis server
and used by the \pilot-Manager to conduct e.\,g.\ placement decisions; (ii) \cu
are stored in several queues. Each \pilot-Agent generally pulls from two
queues: its agent-specific queue and a global queue. Since the Redis server is
globally available, it also serves as central repository that enables the
seamless usage of BigJob from distributed locations. That means that
application can easily re-connect to a \pilot and \computeunit,
via a unique URL.

{\it Data Management:} BigJob supports two forms of data management: (i) in the 
push-based mode all data transfer are handled by the \pilot manager, (ii) in 
the pull mode the \pilot-Agent downloads the data before running a 
\computeunit. Further, there are two types of data: (i) data associated with a 
\pilot and (ii) data associated with a \computeunit. For each \pilot instance a 
sandbox is created; every \computeunit is assigned a directory in this sandbox. 
For every \computeunit both \pilot and \computeunit data files are made 
available in sandbox of the \computeunit and can be accessed by the application 
via their I/O subsystem (e.\,g.\ the Posix API or a specialized I/O 
library, such as HDF5).

{\it Fault Tolerance:} Ensuring fault tolerance in distributed environments is 
a challenging task~\cite{Fischer:1985:IDC:3149.214121}. BigJob is designed to
support a basic level of fault tolerance. Failures can occur on many levels: on
hardware, network, and software level. The complete state of BigJob is
maintained in the distributed coordination service Redis, which stores the
state both in-memory and on the filesystem to ensure durability and
recoverability. Both the application and the \pilot-Manager can disconnect from
running \pilot-Agent and re-connect later using the state within Redis. Also,
the agent and manager are able to survive transient Redis failures. To address
permanent Redis failures additional pre-cautions are required, e.\,g.\ a
redundant Redis server setup with failover. Nevertheless, in most cases the
ability to quickly restart the Redis server (if necessary on another resource)
is sufficient. Another error source are file movements: \pilotdata currently
relies on the built-in reliability features of the transfer service; Globus
Online e.\,g.\ automatically restarts failed transfers. In the future, we will
provide fault tolerance also for non-benign faults, e.\,g.\ network partitions,
resource slowdowns etc.

\subsection{Pilot-API: An Abstraction for Distributed Data \& Compute}
\label{sec:pilot-api}

The Pilot-API~\cite{pilot_api} is an \emph{implementation abstraction}
providing a well-defined control and programming interface to \pilotjobs and
\pilotdata. It builds upon earlier work in context of P*~\cite{pstar12} -- a
{\it conceptual} model and abstraction that enabled the reasoning about
\pilotjobs and \pilotdata in a semantically consistent way. The \pilot-API is
designed to be interoperable and extensible and exposes the core
functionalities of a \pilot framework and can be used across multiple distinct
production cyberinfrastructures.

\begin{figure}[t]
	\centering
		\includegraphics[width=0.5\textwidth]{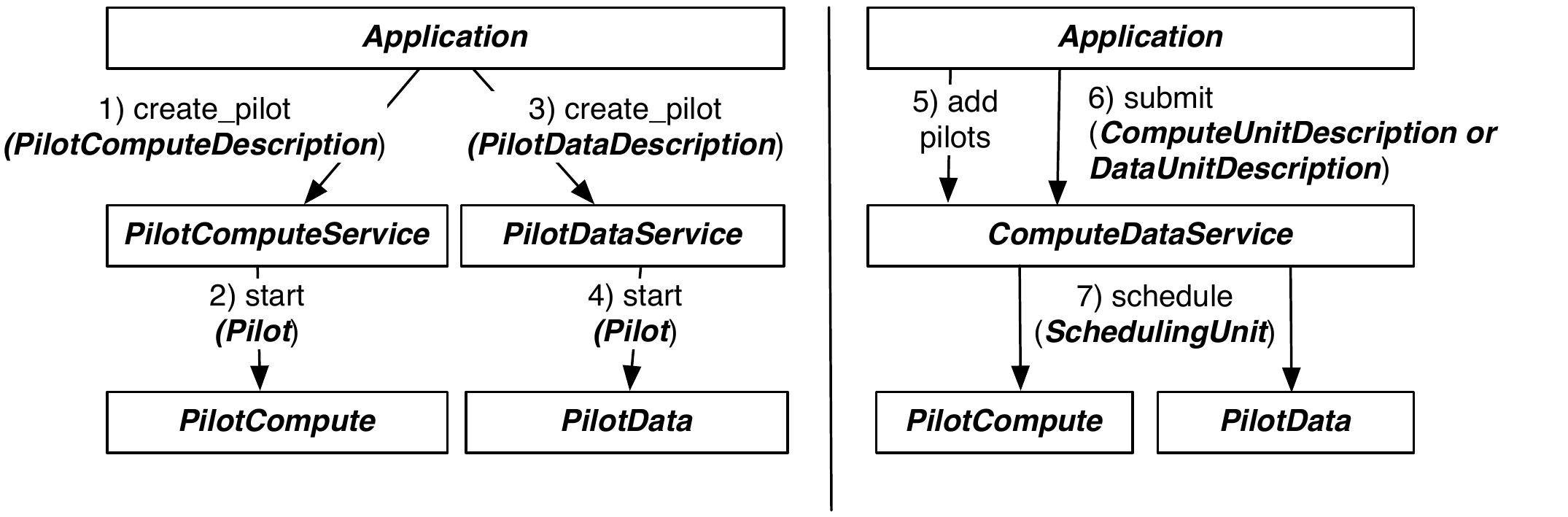}
                \caption{\footnotesize{The Pilot-API
                    exposes two primary functionalities: The
                    PilotComputeService and PilotDataService are used
                    for the management of \pilotcomputes and
                    \pilotdata.  The application workload is submitted
                    via the \computedataservice.\upp}}
	\label{fig:figures_pilot-api}
\end{figure}

Defining the right abstraction for managing computational, data-intensive tasks
and distributed resources is a challenging task and requires trading-off
contradictory objectives, such as simplicity vs. flexibility etc. simplicity.
One of the best-known resource management abstraction is a \emph{process}. A
process encapsulates an executing program providing the abstraction of a
virtual CPU and memory~\cite{Bell_timesharing}. While the illusion of infinite
resources simplifies application development, distributed environments are
generally too complex to maintain this abstraction. Thus, in a distributed
environment a program commonly consists of multiple processes. Programs are
executed via a resource manager using the job abstraction. A \emph{job} denotes
to the batch execution of a program without user intervention on a set of
(possibly distributed) resources. In contrast to local processes, an
application typically is required to specify the resources requirements,
i.\,e.\ the number of cores/resource slots, memory etc. Commonly, the job
abstraction is used for managing compute and data-intensive applications in HPC
and HTC environments. It also provides the basis for the SAGA job
model~\cite{sagastuff}. The \pilot-API relies on a similar model; however, it
separates resource allocation (i.\,e.\ the start of the \pilot) from the actual
execution of the workload providing applications with the ability to use
late-binding when assigning compute/data to resources. This approach is also
referred to as multi-level scheduling. To support multi-level scheduling the
\pilot-API provides two packages: (i) one for resource allocation and \pilot
management and (ii) one for application workload management (see
Figure~\ref{fig:figures_pilot-api}). In this section we describe the
fundamental abstractions and usage models of both parts of the \pilot-API.

\subsubsection{Resource Allocation and Pilot Management} 

The first part of the \pilot-API is concerned with the management of  the 
lifecycle of \pilots, i.\,e. \pilotcomputes and \pilotdata. A
\pilotcompute allocates a set of computational resources (e.\,g.\ cores). A
\pilotdata is conceptually similar and represents a physical storage resource 
that is used as a logical container for dynamic data placement, e.\,g.\ for 
compute-local data replicas or for caching intermediate data.

A \pilotcompute marshals the the job running the \pilot-Agent; it is
responsible for managing a set of resource slots acquired from the local
resource manager. The instantiation of \pilotcomputes are done via a factory
class, the \pilotcomputeservice, using a description object containing the
resource requirements of the application, the \pilotcomputedescription. The
description comprises of a service URL referring the resource manager used for
instantiating the \pilot, a process count specifying the number of required
resource slots and several optional (potentially backend-specific) attributes.
Further, the \pilotcompute API provides methods for managing the lifecycle of
the agent job, i.\,e.\ for querying its current state and runtime attributes
and for canceling it.

Similarly, \pilotdata objects are created via the \pilotdataservice class. A
\pilotdata refers to a physical storage location, e.\,g.\ a directory on a
local or remote filesystem or a bucket in a cloud storage service. Similarly, 
lifecycle methods for the management and querying of the storage resource are 
provided.

Using \pilot-Abstractions different types of distributed compute resources,
storage infrastructures and transport protocols can be marshaled into an
application-specific resource overlay. Once an application has started a set of 
\pilotcomputes and \pilotdata -- the control of this resources is delegated to 
the application, which can then utilize them accordingly and optimize execution 
with respect to computational/memory requirements of its tasks and/or data 
locality.

\subsubsection{Application Workload Management} 
\label{sec:pilot-api-workload-management}
The \pilot-API provides the \dataunits (\du) and \computeunit (\cu) classes as
the primary abstraction for expressing and managing application workloads.
Using these two primitives, applications can specify computational tasks
including their input and output files. A \cu represents a self-contained piece
of work, while a \du represents a self-contained, related set of data. A \cu
encapsulates an application task, i.\,e.\ a certain executable to be executed
with a set of parameters and input files. A \du is defined as an
\texttt{immutable} container for a logical group of ``affine'' data files,
e.\,g.\ data that is often accessed together e.\,g.\ by multiple \computeunits.
This simplifies distributed data management tasks, such as data placement,
replication and/or partitioning of data, abstracting low-level details, such as
storage service specific access details. A \du is completely decoupled from its
physical location and can be stored in different kinds of backends, e.\,g.\ on
a parallel filesystem, cloud storage or in-memory. Replicas of a \du can reside
in different \pilotdata.

A \computeunit is a computational task that operates on a set of input data
represented by one or more \dataunits. Further, \dataunits can be bound to a
\pilotcompute facilitating the reuse of data between a set of \computeunits,
e.\,g.\ to efficiently support iterative applications. The output of a
\computeunit can be written to a set of \dataunits. The runtime system ensures
that the logical references to a \du will be resolved and ensures that the
files are made available in the sandbox of the \cu, i.\,e.\ if necessary the
files corresponding to the \du will be moved. Using these two core abstractions
for application workloads an application can compose complex application
scenarios consistent of multiple \computeunits and \dataunits. Both \cus and 
\dus are described by the use of a \computeunit-Description
(CUDs) and a \dataunit-Description (DUDs) objects defined in the JSON format. A
\dud contains all references to the input files that should be used to
initially populate the \du.

\begin{figure}[t]
	\centering
	  \includegraphics[width=0.49\textwidth]{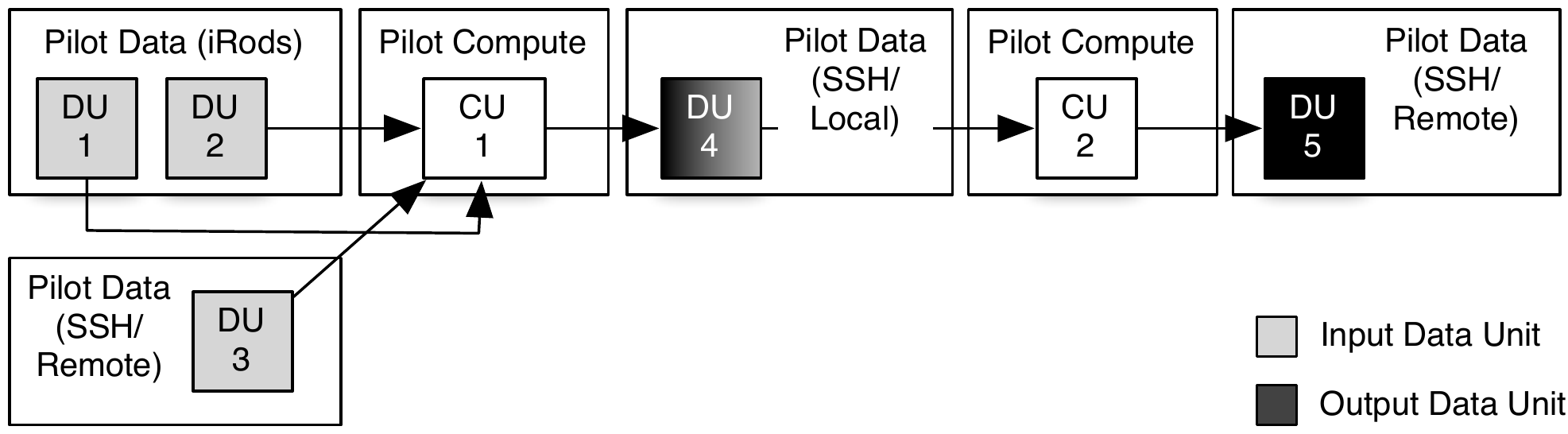}
          \caption{\textbf{DU and CU Interactions and Data Flow:} Each
            \cu can specify a set of input \du. The framework will
            ensure that the \du is transferred to the \cu. Having
            terminated the job, the specified output is moved to the
            output \du.\upp\upp\upp}
	\label{fig:figures_data-flow}
\end{figure}

Further, applications can express data/compute dependencies on an
abstract, high level using \pilot-API. Figure~\ref{fig:figures_data-flow} shows an example of a
data flow between multiple phases of compute. As described, applications are
required to organize their data in form of \dus, which represents a logical
group of files. A \du can be potentially placed in multiple \pilots to
facilitate fault tolerance or a faster access. Applications can declaratively
specify \cus and \dus and effectively manage the data flow between them using
the \pilot-API. A \cu can have input and output dependencies to a set of \dus.
For this purpose, the API declares two fields in the \computeunitdescription:
\texttt{input\_data} and \texttt{output\_data} that can be populated with a
reference to a \du. The runtime system ensures that these dependencies are met
when the \cu is executed, i.\,e.\ either the \dus are moved to a \pilot that is
close to the \cu or the CU is executed in a \pilot close to the \du's \pilot.
In the best case, the \pilotdata of the dependent \dus is co-located on the
same resource as the \cu, i.\,e.\ the data can be directly accessed via a
logical filesystem link. Otherwise, the data is moved via a remote transfer.
Further, a \cu can constrain its execution location to a certain resource. The
input data is made available in the working directory of the \cu.

The \pilot-API provides various levels of control on how the application
workload is managed: (i) applications can either bind their workload (i.\,e.\
\cus and \dus) directly to a \pilot (a \pilotcompute or a \pilotdata) using
their own application-level scheduling mechanisms or (ii) applications can
utilize a workload management service, such as the \computedataservice
introduced in the next section. \cu and \du descriptions are submitted to one 
of these services, which returns a \computeunit/\dataunit instance. This 
instance can then be used for state queries and lifecycle management (e.\,g.\ 
canceling a \cu).

\section{\computedataservice: A Workload Management Services based on Affinities}  
\label{sec:pilot-abstraction-scheduling}
\label{sec:pilot_scheduling}
\label{sec:affinity}

Typically network bandwidth within cluster and even more in WAN settings are
oversubscribed by a significant factor, ignoring the locality thus, can have a
severe impact on an application's performance. Different investigations
(e.\,g.\ \cite{ddia_ptrsa10, 1029935}) have shown that considering data/compute
entities equally while making placement decisions leads to performance gains.
An important consequence of data and computation as equal first-class entities
is that either data can be provisioned where computation is scheduled to take
place (as is done traditionally), or compute can be provisioned where data
resides. This equal assignment of \computeunits leads to a richer set of
possible correlations between the involved \dus and \cus; correlations can be
either spatial and/or temporal. These correlations arise either as a
consequence of constraints of localization (e.\,g.\ data is fixed, compute must
move, or vice-versa), or as temporal ordering imposed on the different
\dataunits and \computeunits.

\begin{figure}[t]
	\upp
	\centering
		\includegraphics[width=0.5\textwidth]{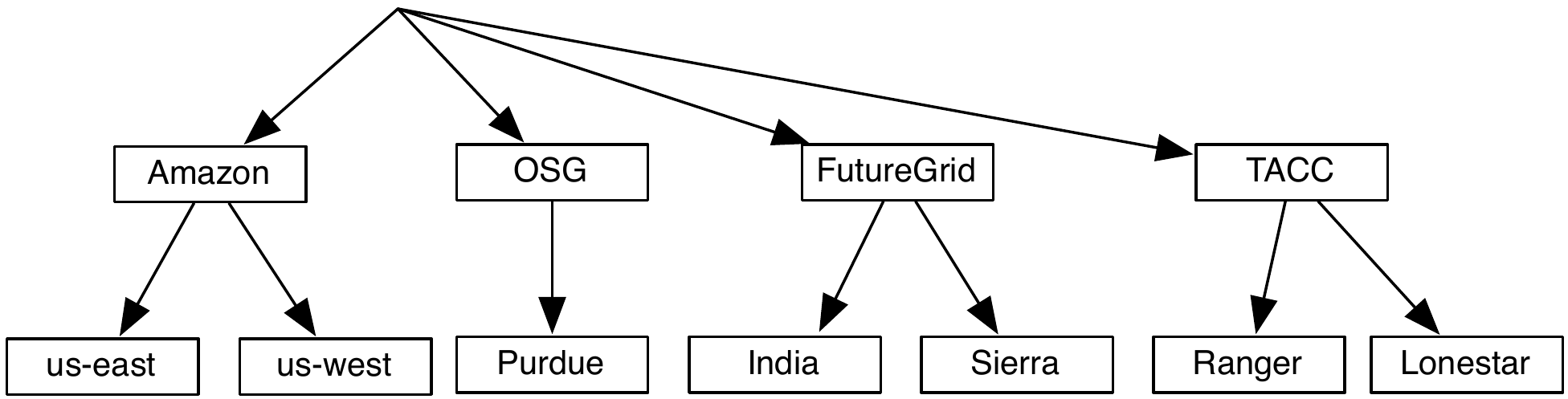}
	\caption{\textbf{Affinities between Distributed Resources:} 
	\pilotdata assigns each resource an affinity based on a simple 
	hierarchical model. The smaller the distance between two resources, the 
	larger the affinity.\upp\upp\upp}
	\label{fig:figures_affinities}
\end{figure}

\noindent
\emph{Resource affinities} describe the relationship between a set of compute
and/or storage resources. We use a simple model for describing resource
affinities: data centers and machines are organized in a logical topology tree.
The further the distance between two resources, the smaller their affinity.
Figure~\ref{fig:figures_affinities} shows e.\,g.\ how a distributed system
consisting of different types of cloud and grid resources can be modeled. Using
such an resource topology, the runtime system can deduce the connectivity
between two resources, to estimate e.\,g.\ the costs induced by a potential
data transfer. While this model is currently very coarse grained, it can be
enhanced by assigning weights to each edge to reflect dynamical changes in
factors that contribute to connectivity. The affinity of a \pilot is determined
based on the resource it is located on, i.\,e.\ the proximity of two \pilots is
deduced from the distance of their resource in the resource topology tree. The
mapping between resource and \pilot is done by assigning each \pilot a logical
location using a user-defined affinity label in the \pilotdescription. This
logical location assignment is utilized by the scheduler to create the resource
topology tree.

\noindent
\emph{Compute/data affinities} describe the relationships between \dus and
\cus. A \du e.\,g.\ is the primary abstraction for the logical grouping of
data. \cus can have input and output dependencies to a set of \dus, i.\,e.\
the data of these \dus is required for the computational phase of the \cu. The
output data is automatically written to one or more output \dus. The framework
utilizes these affinities to place \dus and \cus into a suitable \pilotcompute
or \pilotdata. Further, \cus and \dus can constrain their execution resource
to a particular affinity (e.\,g.\ to a certain location or sub-tree in the
logical resource topology). The runtime system then ensures that
the data and compute affinity requirements of the \cu/\du are met.

\noindent
\emph{\pilot-based Scheduling:} 
BigJob provides a rudimentary but an important proof-of-concept affinity-aware
scheduler that attempts to minimize data movements by co-locating affine \cus
and \dus to \pilots with a close proximity. The scheduler is a plug-able
component of the runtime system and can be replaced if desired. The default
implementation relies on the resource topology and affinity attributes
provided via the \pilot-API to reason about the relationships between \dus,
\cus, \pilots and resources to optimize data localities and movements. The
affinity-aware scheduler
currently implements a simple strategy based on earlier
research~\cite{ddia_ptrsa10} that suggests that considering both data
and compute during placement decisions leads to a better
performance. As shown in
Figure~\ref{fig:figures_computedataservice-scheduling}, BJ relies on
two queues for managing \cus. \cus without any affinity are assigned
to the global queue from where they can be pulled from multiple
\pilot-Agents. If there is affinity to a certain \pilot because the
input data resides in this \pilotdata, the \cu can be placed in a
\pilot specific queue. For each \cu the following steps are executed:
\begin{compactenum}
	\item The \pilot-Manager attempts to find a \pilot that best fulfills the 
	requirements of the \cu with respect to (i) the requested affinity and  
	(ii) the location of the input data.
	\item If a \pilot with the same affinity exists and \pilot has an empty 
	slot, the \cu is placed in this pilots queue.
	\item If delayed scheduling is active, wait for $n$ sec and recheck 
	whether \pilot has a free slot.
	\item If no \pilot is found, the \cu is placed in global queue and pulled 
	by first \pilot which has an available slot. 
\end{compactenum}
The \pilot-Agent that pulls the \cu from a queue is responsible for
ensuring that the input \du is staged to the correct location, i.\,e.\
before the \cu is run, the \du is made available in  
the working directory of the \cu either via remote transfer or a logical link.

\section{Experiments}
\label{sec:experiments}

It is important to appreciate that experiments that aim to
characterize the performance of an abstraction are by their very
nature difficult. We cite two primary reasons: the first is that an
abstraction is only as good as the infrastructure that it is
implemented on. Furthermore, what \pilotdata provides is a uniform way
of reasoning about compute-data distribution and implementing them,
not necessarily new capabilities in and of themselves.  Not
surprisingly, our experiments do not aim to understand the performance
of \pilotdata per se, but the application performance that can be
enabled by the use of Pilot-Data.

Before we discuss experiments in the next sub-section, we develop some
minimal terminology that enables such reasoning across different modes
of distribution and infrastructure, as well as understand the primary
components and trade-offs to determine compute-data placement. We
continue with the description of several experiments aimed at
understanding three different aspects of \pilotdata: (i) In
section~\ref{sec:pd_performance} we demonstrate a proof of existence
and correctness of \pilotdata via the ability to provide uniformity of
access and usage modes for different infrastructures (e.\,g.\ XSEDE
and OSG); (ii) In section~\ref{sec:ngs-experiments} we discuss how
\pilotdata provides a conceptually simple and uniform framework to
reason about how and when to distribute over very different and
architecturally distinct infrastructures, and (iii) In
section~\ref{sec:perf_scalable} we discuss some advanced capabilities
and performance advantages arising from the ability to use \pilotdata
to select ``optimal'' usage modes and support scalability of
large-scale data-intensive applications.

\subsection{Reasoning About Compute-Data Placement}
\label{sec:compute_data_strategies}

A question that arises in the design of systems and that distributed
data-intensive applications have to address, is whether to assign and
move computational tasks to where data resides, or to move data to
where computational tasks can be executed.  An associated question is
when to commit to a given approach. Additionally, if replication is an
option, applications and systems have to determine what the degree of
replication of data should be, and possibly where to replicate.  As an
abstraction for distributed data, \pilotdata must provide the ability
to answer the above questions and implement the results.  To
programmatically determine the best approach and to understand
\pilotdata based experiments, the value of several parameters have to
be considered:

\begin{compactitem}
\item $T_Q$ defined as the queue waiting time at a given resource. 
${T_{Q_{Pilot}}}$ is the queue time of the \pilotjob. ${T_{Q_{Task}}}$ 
is defined as the \pilot-internal queueing time. 
\item ${T_C}$ defined as the compute time.
\item $T_S$ is the staging time, which is defined as the transfer time 
  $T_X$ plus the time to register the data into the system.
\item $T_R(R)$ defined as the time to replicate
  data, 
  where R is the number of sites that data is replicated over.
\item $T_D$ is the time at which data will be accessible across all
  distributed resources.  When replication is involved, it is defined
  as the sum of the $T_R(R)$ and $T_S$.
\end{compactitem}
The relative values of the parameters above, provide the basis to
reason about whether to process/compute where the data already
resides, or whether to move data to where the processing power lies;
they also provide insight on how to possibly distribute data or not.
To a first approximation, which of the two approaches should be
employed is given by the relative values of $T_D$ and the typical
value of ${T_Q}$. The appropriate mode is amongst other things,
strongly dependent upon data volumes in considerations, as well as the
capability of the tools and middleware in use.

When both data and compute can be scheduled/placed, the decision about
which entity to place/schedule first and which to move -- compute to
data or data to compute -- is determined via a simple trade-off
between ${T_Q}$ and $T_X$.  
If the expected $T_X$ is larger than the $T_Q$, then the compute is assigned
to a site first, and subsequently data is placed. 
When data is already distributed, compute resources have to
be chosen in response to this distribution. Resources co-located with data
replicas, with the lowest queue waiting time present optimal choice. However,
it is important to appreciate that there is an overhead in ensuring that data
is replicated in a distributed fashion.

In practice hybrid modes can be employed. As an example, distributed
data replication can initially be set to be partial, viz., only over a
subset of possible distributed sites. In other words, replication
might commence over a subset of suitably chosen nodes, followed by a
sequential increase in the replication (factor) if compute resources
close to the replica do not have sufficient compute capacity. Currently these 
decisions are made manually, but eventually such
scheduling and placement decisions will be driven by both application
and system information; the affinity model discussed provides one way
of providing this information to the ``scheduling engine''.

As with any model, it is important to recognize its practical limitations. In
many cases applications content for shared resources, such as the network or a
shared storage systems. Thus, $T_X$ e.\,g.\ will dependent on additional
external factors (such as the current network utilization).

\subsection{Understanding Pilot-Data}
\label{sec:pd_performance}

The objective of the first set of experiments is to demonstrate the ability of
\pilotdata to marshal different storage backend infrastructures. We then
characterize the performance of \pilotdata on different cyberinfrastructures
(e.\,g.\ XSEDE and OSG) by investigating the different components of the data
distribution time $T_D$; some of these
experiments involve investigating the impact of replication.  For
joint submission machine to XSEDE and OSG resources, we utilize GW68
-- a gateway node located at Indiana University and part of the XSEDE
infrastructure.

\begin{figure}[t]
	\upp
	\centering
	\includegraphics[width=0.45\textwidth]{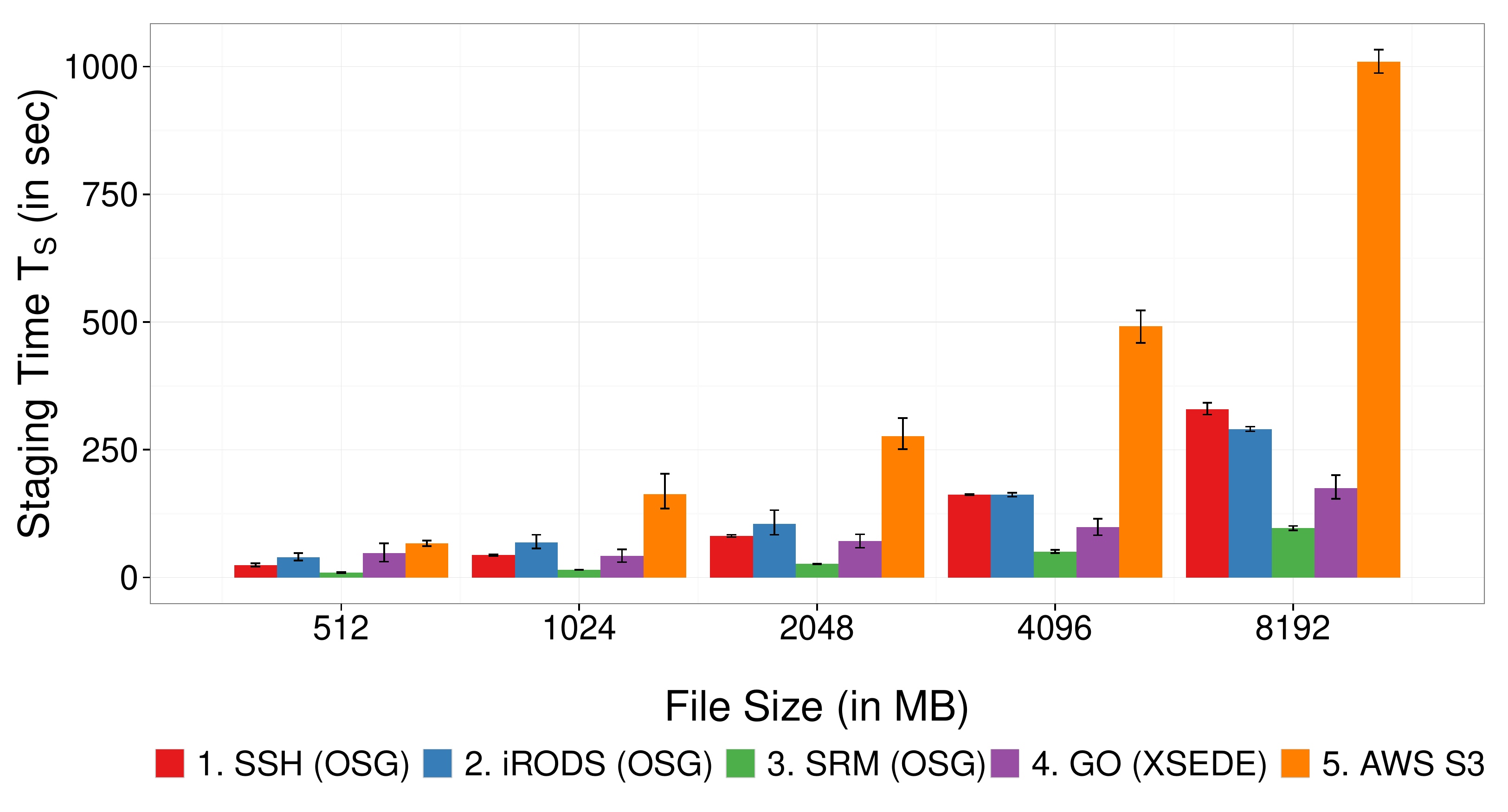}
	\caption{\textbf{\pilotdata on Different Infrastructures:}
          Time to instantiate a \pilotdata with a dataset of given
          size. For \irods, we measure only the staging time and not the
          time required to replicate the data. SRM performs best mainly due
		  to the reliance on GridFTP as data transfer protocol. SSH and \irods 
		  show an acceptable performance for smaller datasets. Globus Online is
		  associated with some overheads due to its service-based nature, which 
		  is particularly visible for smaller data sizes. S3 is constrained by the
		  limited bandwidth available to the Amazon datacenter.
		  }
	\label{fig:performance_irods_irods-ssh}
\end{figure}

In the first experiment we investigate $T_S$ on different \pilotdata backends.
Figure~\ref{fig:performance_irods_irods-ssh} illustrates $T_{S}$ for
different backends, i.\,e.\ the time necessary to populate a
\pilotdata on different infrastructures: in scenario 1 the PD is mapped to a
directory on an OSG submission machine, in scenario 2 on an \irods
collection on the OSG \irods infrastructure, in scenario 3 on a SRM directory, 
in scenario 4 on a directory on Lonestar accessed via Globus Online and 
in scenario 5 on an Amazon S3 bucket.

The performance primarily depends on the infrastructure used and in particular
the available bandwidth between the submission machine and the storage backend.
$T_S$ is dominated by $T_{X}$, i.\,e.\ the time necessary to transfer files to
the \pilotdata location. Experiments with smaller data sizes have shown that
$T_{register}$ is negligible. Thus, the runtime is directly influenced by the
available bandwidth and the characteristics of the respective transfer
protocol. SRM on OSG clearly shows the best performance: SRM is a highly
optimized storage backend which is in this scenario used with GridFTP a highly
efficient data transfer protocol. Globus Online particularly performs well for
larger data volumes: the service also utilizes GridFTP however adds an
additional management layer. For smaller data volumes SSH is a better choice.
The initialization for setting up an SSH connection is significantly lower than
for the creation of a Globus Online request. With larger data volumes the
initialization overhead becomes insignificant and Globus Online benefits from
the more efficient GridFTP transfer protocol. $T_S$ for \irods behaves
comparable to $T_S$ for SSH. $T_{S}$ for S3 increases linearly -- an indicator
that the bandwidth to the AWS site is a limiting factor. It is noteworthy that
\pilotdata can support various combinations of different storage and transfer
protocols giving the application the possibility to chose an appropriate
backend with respect to its requirements; e.\,g.\ while Globus Online is the
best choice for large volume transfers within XSEDE, data-intensive
applications are required to use \irods.

In the second experiment, we evaluate the impact of replication ($T_R$) on
$T_D$. If system-level support for replication is provided, e.\,g.\ by a
distributed data management middleware such as \irods, \pilotdata can utilize
this capability as a dynamic caching mechanism (to be contrasted with the
usage of \irods for data storage and management).
In the following experiment, we investigate $T_R$ for different
infrastructures and configurations: (i) \irods/OSG with group-based
replication (osgGridFTPGroup), (ii) \irods/OSG with sequential
replication in which one replica is created after the other and (iii) SRM with
sequential replication.

\begin{figure}[t]
 \centering
 \includegraphics[width=0.45\textwidth]{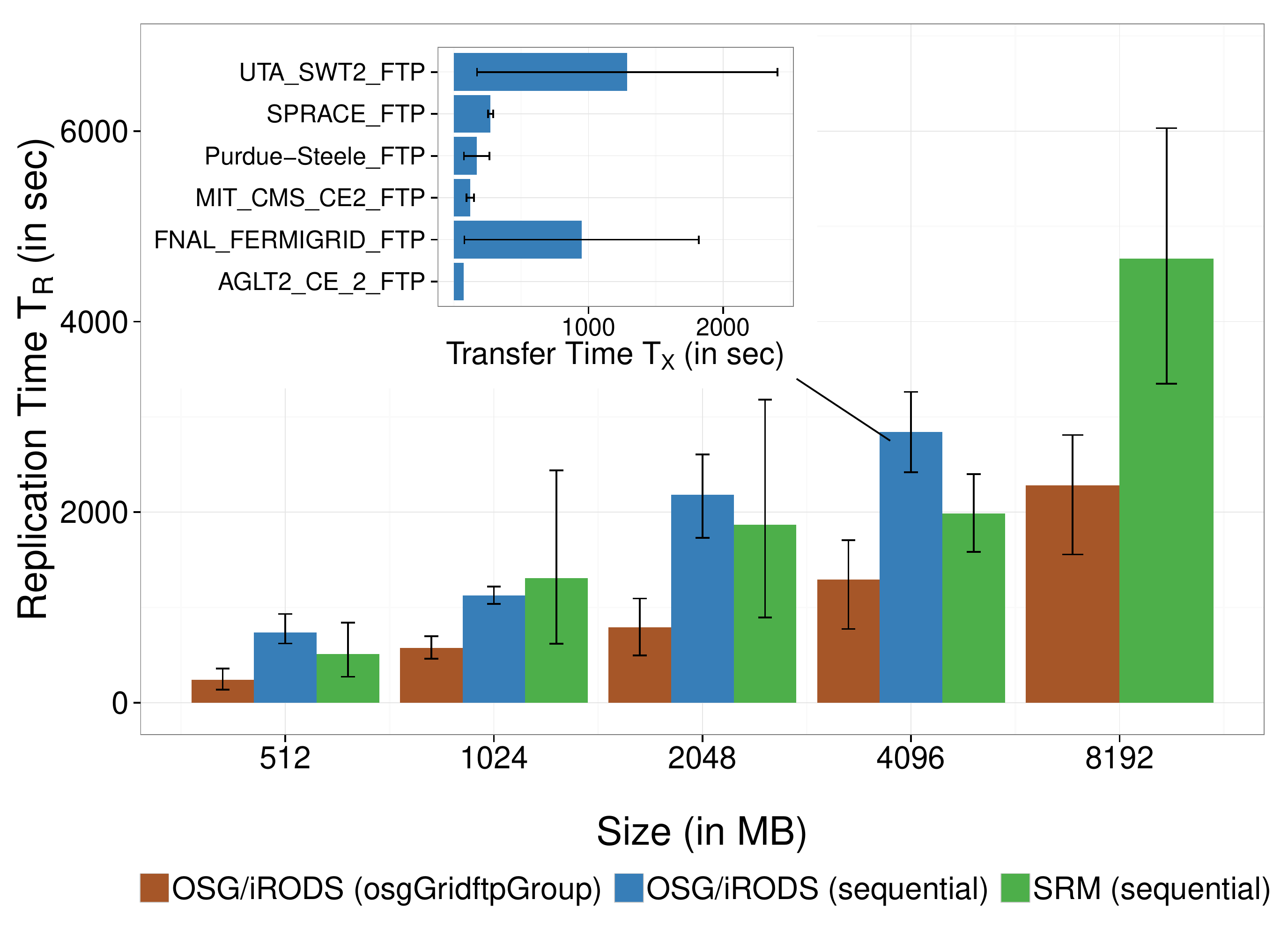}
 \caption{\textbf{Using Replication on OSG:} $T_R$ on OSG: in scenario 
   osgGridFTPGroup data input to 9 \irods resources that are members of this 
   group, in the sequential scenario 6\,\irods respectively SRM resources are 
   used. The inset shows the distribution of $T_R$ with respect to the 
   different hosts for the 4\,GB \& OSG/\irods scenario.   }
  	\label{fig:performance_irods_irods}
\end{figure}

Figure~\ref{fig:performance_irods_irods} illustrates the results: On OSG the
$T_R$ for the group-based replication both with \irods as well as with SRM is
significantly better than for the sequential replication. In case of sequential
replication SRM performs better than \irods, i.\,e.\ while \irods adds the
ability to manage the data distribution on OSG on a higher level, it also adds
some overhead. In both cases, the frequency of failures was very high. While
the \texttt{osgGridFtpGroup} group consisted of 9 nodes, the average number of
resources that actually received a replica was $\sim$7.5. In general, the
overall performance is determined by the available bandwidth between the
central \irods server (located at Fermilab near Chicago) and the individual
sites. For 4\,GB case in scenario (ii), the individual $T_X$ are depicted in
the inset of Figure~\ref{fig:performance_irods_irods}. 

OSG provides the user with a variety of storage services and thus, options to
organize their compute and data. Abstraction such as \pilotdata are important
to provide a unified access to these service and to enable the application to
reason about the distribution of their compute and data. While e.\,g.\
sequential replication is well suited for creating a small number of replicas,
it is only beneficial a small amount of additional compute resources for a
dataset is required. In other cases, e.\,g.\ in order to supporting larger
amounts of compute, an OSG wide replication using \irods is beneficial. Another
important observation is the fact that different sites have very different
performance characteristics. Thus, the ability for applications to optimize
data/compute placements with respect to their computational and data
requirement presents both plenty of opportunities but also is somewhat
challenging. \pilotdata provides a unified interface which allows applications
to trade-off different infrastructure capabilities and characteristics to
enable an efficient execution of its workload.

\subsection{Understanding Pilot-Abstractions on Heterogeneous Infrastructure}
\label{sec:ngs-experiments}

Infrastructures significantly differ in the way they manage data and
compute; e.\,g., on XSEDE resources it is generally possible
to place data on the distributed filesystem mounted to all compute
nodes. On OSG this is not simply possible
since users generally cannot access compute nodes without Condor;
however, the \irods service on OSG enables the application to push
data to the different OSG resources. These different kinds of
semantics increase the complexity for applications to deal with data.
Experiments in this section show how \pilotdata provides a unified, logical
resource abstraction, which allows applications to reason about trade-offs and
pursue different compute/data placements strategies as laid out before, e.\,g.\
bringing compute to data versus data to compute.

For this purpose we use the \pilot-API to manage the input/output data in
conjunction with the computational tasks of the BWA genome sequencing
application~\cite{Li:2010:FAL:1741823.1741825}. The application requires two
kinds of input data: (i) the reference genome and index files ($\sim$8\,GB), 
and (ii) the short read file(s) obtained from the sequencing
machines. The alignment process can be parallelized by partitioning the read
files and processing them using multiple BWA tasks. The reference genome and 
index files are shared between all tasks. The experimental
configuration consists of 2\,GB read files, which are partitioned to 8 tasks
each processing 256\,MB.

In scenarios 1 and 2 we conduct baseline experiments using {\it simple} data
management, i.\,e.\ each task pulls in all input data from the submission
machine (GW68). Tasks are distributed across 8\,\pilots on OSG (scenario 1) and
across a single \pilot marshaling 24 cores on Lonestar/XSEDE (scenario 2); note
that in HTC environments such as OSG, a \pilot typically marshals only a single
core (up to a maximum of one entire node). We restrict OSG resources to a set
of 9 machines, which are supported by the OSG \irods installation. The
resources are distributed across the eastern and central US including resources
at TACC, Purdue and Cornell.  The OSG \pilotcomputes are submitted using the
SAGA-Python Condor adaptor~\cite{saga-python-pd} and GlideinWMS~\cite{5171374},
a workload management system built on top of the \pilot capabilities of
Condor-G/Glide-in~\cite{condor-g}.

In scenarios 3-5 we use the ability to co-locate \pilotcomputes and
\pilotdata. For this purpose, the \du containing the input data is
placed in a \pilotdata close to the \pilotcompute.
In scenario 3, the data is placed and replicated into an \irods-based
\pd (9 machines); in scenario 4 an SSH \pilotdata on the shared Lustre
scratch filesystem of Lonestar is used. In both scenarios the
\pilotcomputes and \pilotdata are co-located.  Finally, we investigate
the ability to use \pilotcomputes and \pilotdata across multiple OSG
and XSEDE resources in scenario 5. In this scenario, the input data
set resides on a \pilotdata on Lonestar.  Two \pilotcomputes are used; One 
\pilotcompute allocating one node with 12\,cores is submitted to Lonestar and 
four \pilotcomputes are spawned on OSG.

\begin{figure}[t]
	\centering 
	\includegraphics[width=0.5\textwidth]{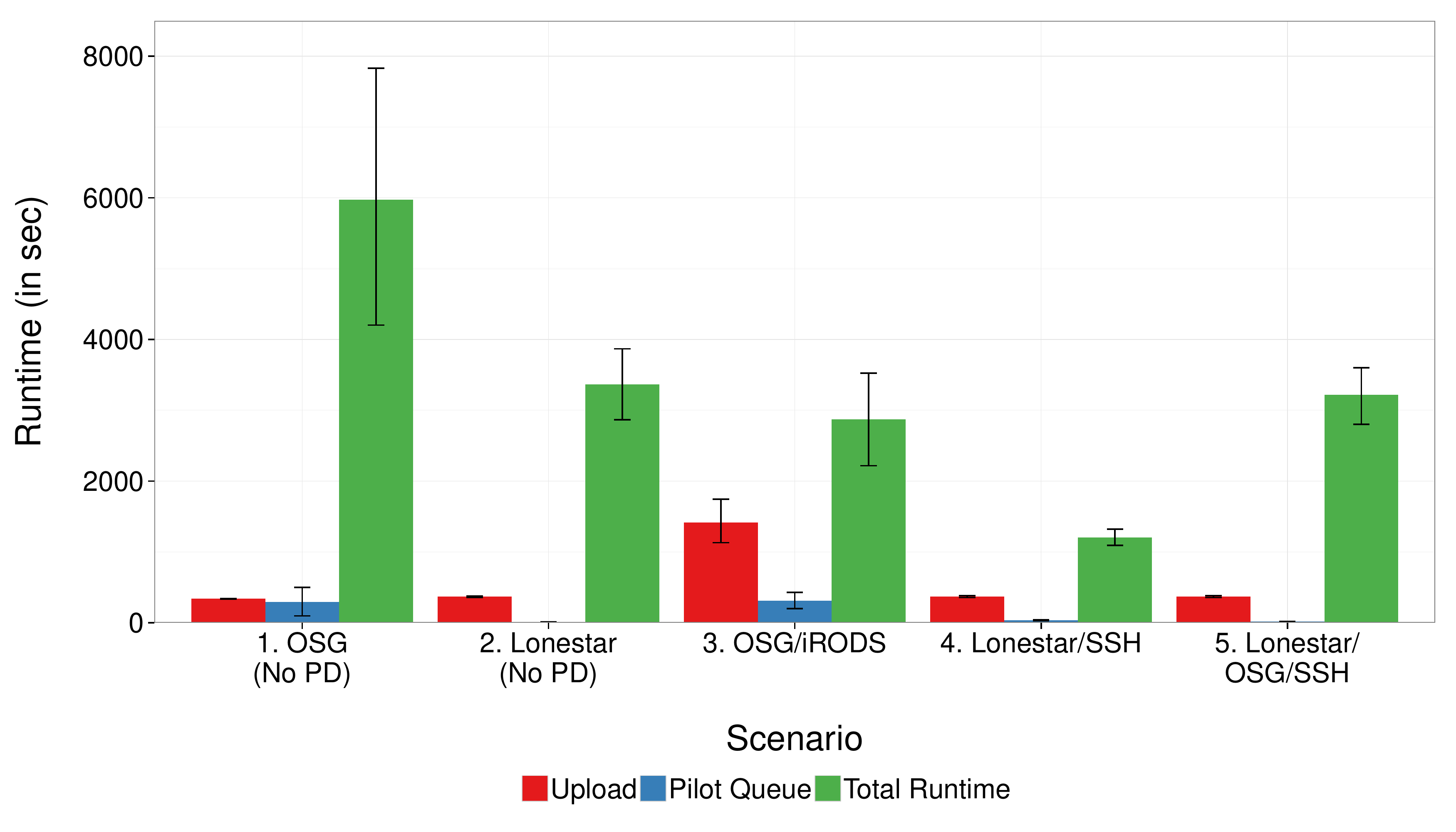}
	\caption{\textbf{Genome Sequencing Using \pilotdata on Different Infrastructures: }
		  Runtimes for running BWA on 2\,GB of
          sequence read files using 8\,tasks for five different
          infrastructure configurations on XSEDE and OSG.  The usage of \pd 
		  (scenarios 3-5) leads to an performance improvement compared to a 
		  naive data management (scenario 1-2).
		  }        
	\label{fig:performance_multi-pilot_multi-pilot}
\end{figure}

\begin{figure}[t]
	\centering
		\includegraphics[width=.5\textwidth]{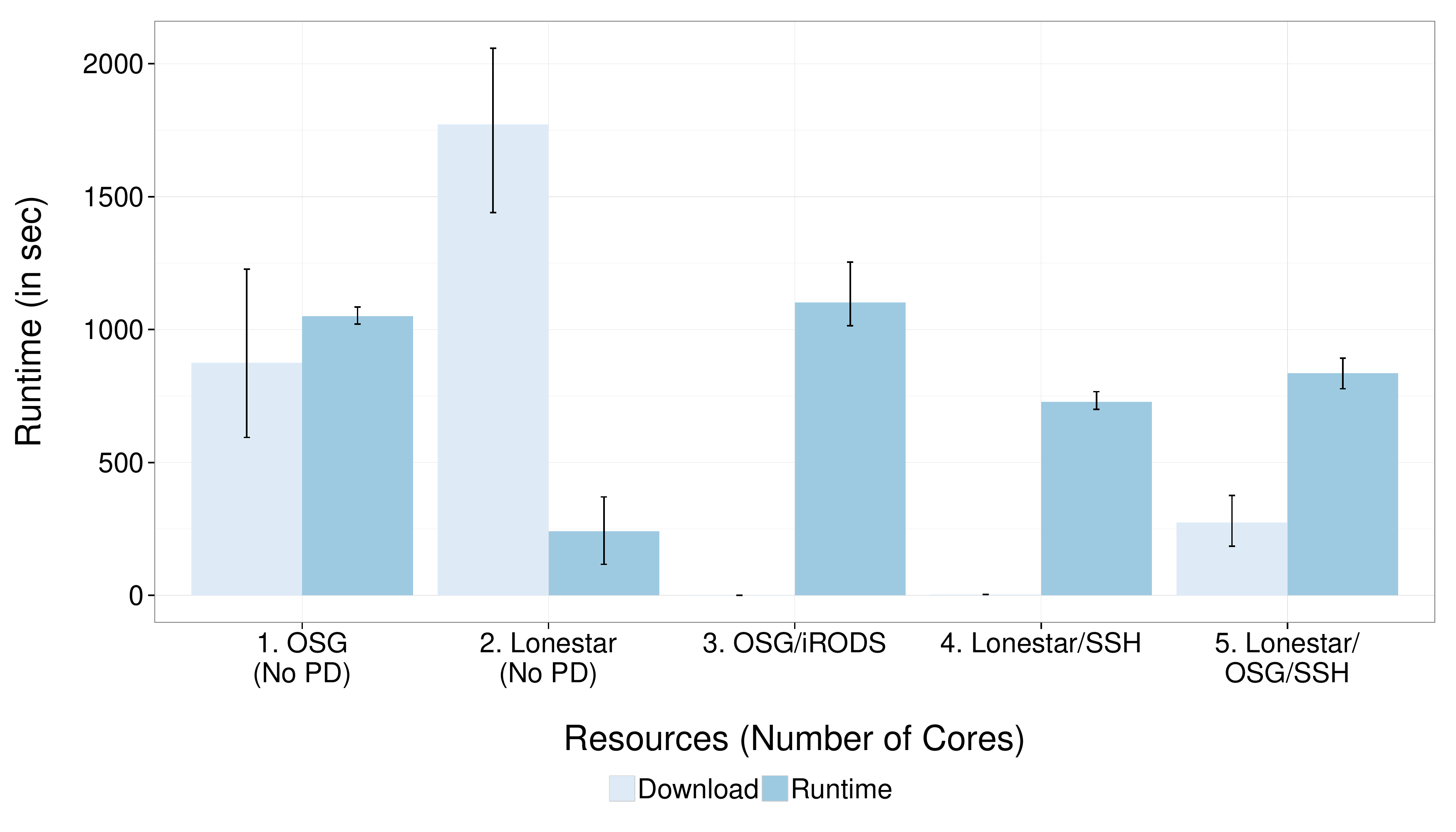}
	\caption{\textbf{Staging and Task Runtimes:} When comparing the individual 
	task staging and runtimes, it becomes obvious that file staging quickly 
	becomes a bottleneck. By using \pilotdata the file staging time (Download) 
	can be significantly reduced. In scenario 5 half of the tasks are required 
	to download the files, thus, a small file staging time remains.}
	\label{fig:performance_multi-pilot_multi-pilot-distributed-cu-runtimes-sep}
\end{figure}

Figure~\ref{fig:performance_multi-pilot_multi-pilot}
and~\ref{fig:performance_multi-pilot_multi-pilot-distributed-cu-runtimes-sep}
analyze the different scenarios. The insert describes $T_{D}$, i.\,e.\ the time
for uploading, inserting and in case of \irods replicating 8.3\,GB of input
data. In general, the \pilot queuing times, i.\,e.\ the time until a \pilot
becomes active, are higher on OSG than on XSEDE resources. The queueing time
mainly depend on three factors: the current utilization of the resource, the
allocation of the user and the overhead induced by the queuing system.

Scenarios 1 and 2 clearly show the limitations of simple data management
approaches. The necessity for each task to pull in 8.3\,GB data remotely leads
to a bottleneck. In scenario 3 we utilize the data/compute co-location
capabilities of the OSG Condor and \irods installation. The runtime $T$ is
significantly improved compared to scenario 1 mainly due to the elimination of
data transfers. However, the upfront costs for creating the \pd and
replicating the data across OSG are higher -- $T_{D_{\irods}}$ is
$\sim$1,418\,sec, $T_{D_{SSH}}$ is only $\sim$338\,sec for scenario 4/5, but
does not have a replication component (see inset of
Figure~\ref{fig:performance_multi-pilot_multi-pilot}). Thus, $T_{S_{CU}}$ is
significantly higher for SSH than for \irods. However, even after including
$T_D$, the performance of \irods is $~$30\,\% better than the SSH
scenario.

In scenario 5, as the input data resides in a \pd on Lonestar, the
staging time for tasks on Lonestar is significantly reduced. Since the
\pilot queuing time on Lonestar was shorter than on OSG, the majority
of the tasks were executed on Lonestar; on average 4.5 out of the
8\, tasks were run on Lonestar.  Finally, scenario 5 shows that if
sufficient compute resources are available close to the data, it is
beneficial to execute tasks close to the data. In particular this scenario 
demonstrates the power of the \pilotdata abstraction, which enables the 
effective and interoperable use of multiple, heterogeneous infrastructures -- 
OSG and XSEDE -- via a unified API.

\subsection{Understanding Scalability, Distribution \& Replication}
\label{sec:perf_scalable}

In this section we investigate the usage of \pilotdata to manage 
distributed data and compute. Reasons for using distributed resources are 
manifold: often data is pre-distributed or the available resources (cores, I/O) 
on a single machines are not sufficient. For example, while the overall I/O 
throughputs on HPC machines, such as Lonestar
or Stamped, and parallel filesystems, such as Lustre or GPFS, are impressive,
we will show that I/O scaling may be constrained for
data-intensive applications at large scale. Further, many resources provide
significantly less I/O capacity than these flagship machines. Also, actual I/O
speeds are highly dependent on the current utilization of the machine.
\pilotdata provides the ability to overcome some of these constraints by
distributing compute/data to multiple distributed resources, potentially
avoiding situations where disk access speeds are the main constraint in
improving performance.

Offloading tasks to distributed/remote resources is a viable strategy to
minimize bottlenecks, such as high queueing times, insufficient compute or I/O
resources for a certain workload, data placed on distributed resources, etc. As
alluded to in section~\ref{sec:compute_data_strategies} there are different
parameters to consider when distributing compute and/or data. The main barrier
for distributing large-scale data-intensive applications across multiple
resources resources is the necessity to move data. We evaluate four scenarios
using \pilotdata with different data placement strategies (with and without 
up-front data replication) on up to three XSEDE machines. For this purpose, we 
use a larger BWA ensemble consisting of 1024 tasks each processing a read file 
of 1\,GB size on different distributed XSEDE
configurations. In total each task consumes 9\,GB and the ensemble 9,200\,GB of
data. For each tasks two cores are requested.

\begin{figure}[t]
    \upp
	\centering
	\includegraphics[width=0.5\textwidth]{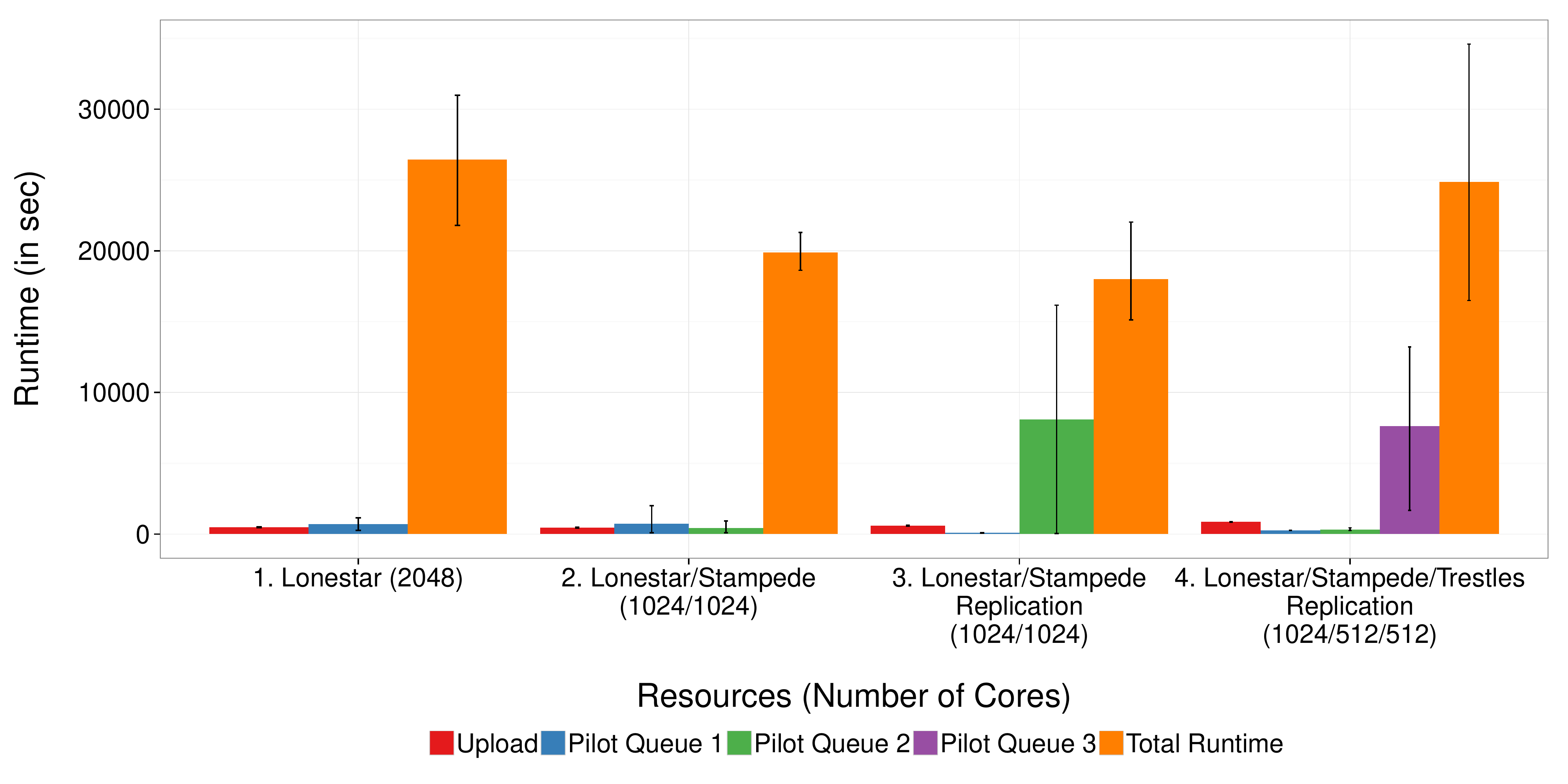}
	\caption{\textbf{Large-Scale, Distributed Genome Sequencing on XSEDE  (Overall Scenario Runtime):} 
	Running 1024 tasks each consuming 9\,GB data on up to three XSEDE machines. 
	The experiment shows that using multiple resources can improve $T$ despite 
	the overhead introduced for data movements. The usage of replication can 
	further improve the runtime $T$ and is essential if the bandwidths to the 
	remote resource is limited as in scenario 4.
}
	\label{fig:performance_scaleout_bwa-1024cus-1GB-distribution}
\end{figure}

\begin{figure}[t]
	\centering
		\includegraphics[width=.5\textwidth]{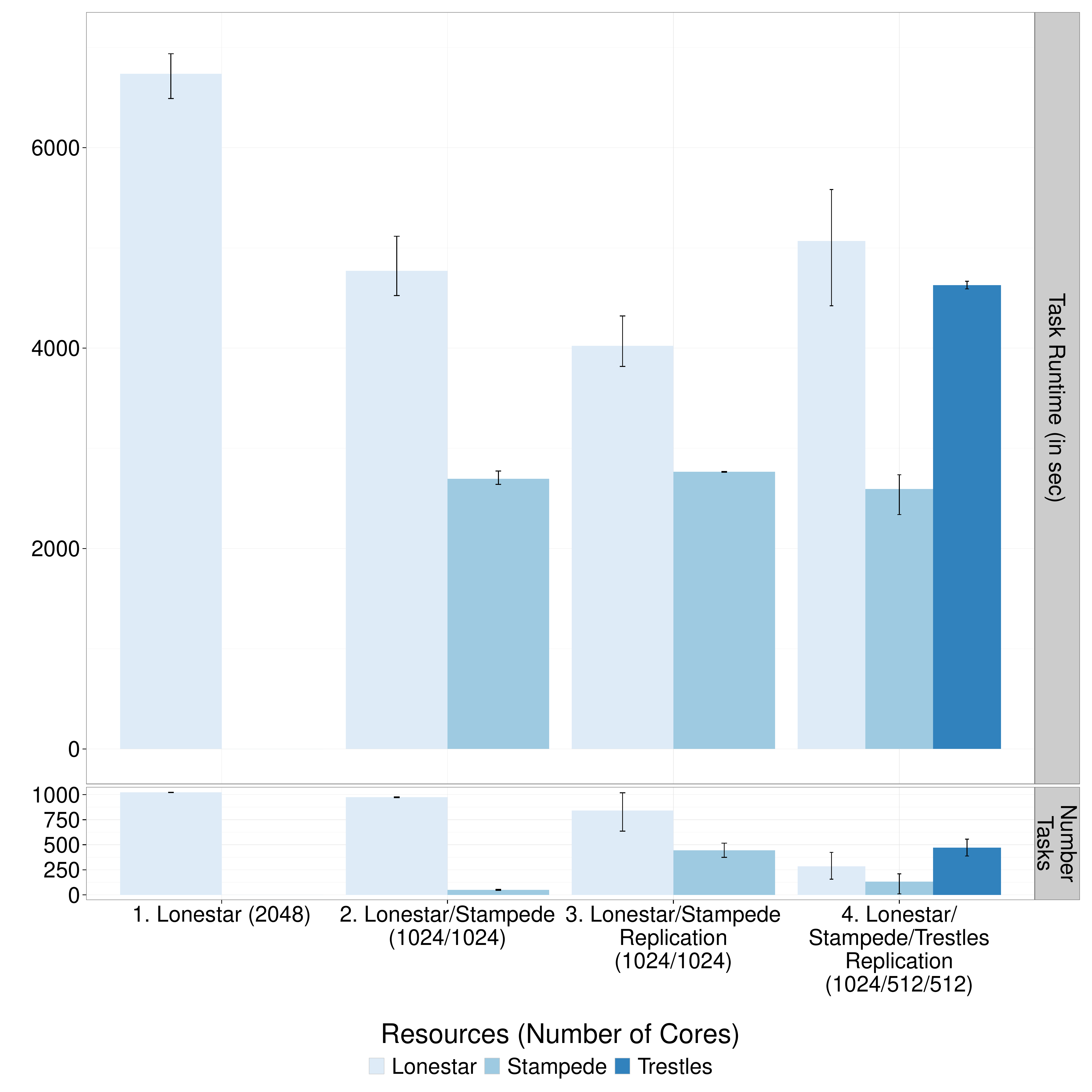}
	\caption{\textbf{Large-Scale, Distributed Genome Sequencing on XSEDE  (Task 
	Runtime and Distribution):} The task runtimes indicate a sensitivity to 
	the number of concurrent tasks on Lonestar, e.\,g.\ when running a majority of the 
	1024 tasks using 2 cores each (scenario 1). 
	In scenario 2 the necessity to move file limits the number of tasks that are executed 
	non-data local on Stampede. The usage of data replication improves the 
	distribution and runtime of the tasks (scenario 3). 
	Figure~\ref{fig:performance_3way-timeseries} analyzes scenario 4. 
	}
	\label{fig:performance_timeseries_cu-runtimes}
\end{figure}

Figures~\ref{fig:performance_scaleout_bwa-1024cus-1GB-distribution}
and~\ref{fig:performance_timeseries_cu-runtimes} summarize the result. As shown
in Figure~\ref{fig:performance_scaleout_bwa-1024cus-1GB-distribution}, the
runtime improves with the number of resources used. Using a single machine, 
such as Lonestar in scenario 1, does not yield in an optimal performance. The 
long runtime of the individual CUs depicted in
Figure~\ref{fig:performance_timeseries_cu-runtimes} indicate a bottleneck on
this machine (very likely the I/O capacity of the Lustre filesystem is
insufficient). Thus, in scenario 2 we distribute the workload to two machines
in close proximity: Lonestar and Stampede (both located at TACC). On each
machine we request a \pilot with 1024\,cores. The overall runtime of
the 1024\,CUs and also the individual CU runtimes improve in this scenario. 
However, the necessity of moving the data lead to another bottleneck: tasks 
that are executed on the remote machine Stampede are required to move 9\,GB of
input data. Moving this data from Lonestar to Stampede required on average 
450\,sec per task. Thus, in total only about 5\,\% of the tasks are executed on
Stampede (see lower part of 
Figure~\ref{fig:performance_timeseries_cu-runtimes}).

To optimize data placements, we deploy data replication in scenario 3: before
the \pilotcomputes and tasks are started a replica of the input \dataunit is
created and placed on Stampede. In average the creation of the replica takes
130\,sec and is negligible in contrast to the overall compute time. A reason
for this is the optimized replication mechanism, which utilizes the replica 
closest to the target site. Thus, an
improvement in $T$ is observable despite the fact that the queuing time on 
Stampede during the time of the experiment was very long (in average 
8100\,sec and thus, about 20 times as long as in 
scenario 2). Considering this, the overall runtime could have been even better 
at a different time. In this scenario, the distribution of the \cus improves; 
despite of the longer queueing time at Stampede about 40\,\% of the tasks are executed on this machine.

\begin{figure}[t]
	\centering
		\includegraphics[width=.5\textwidth]{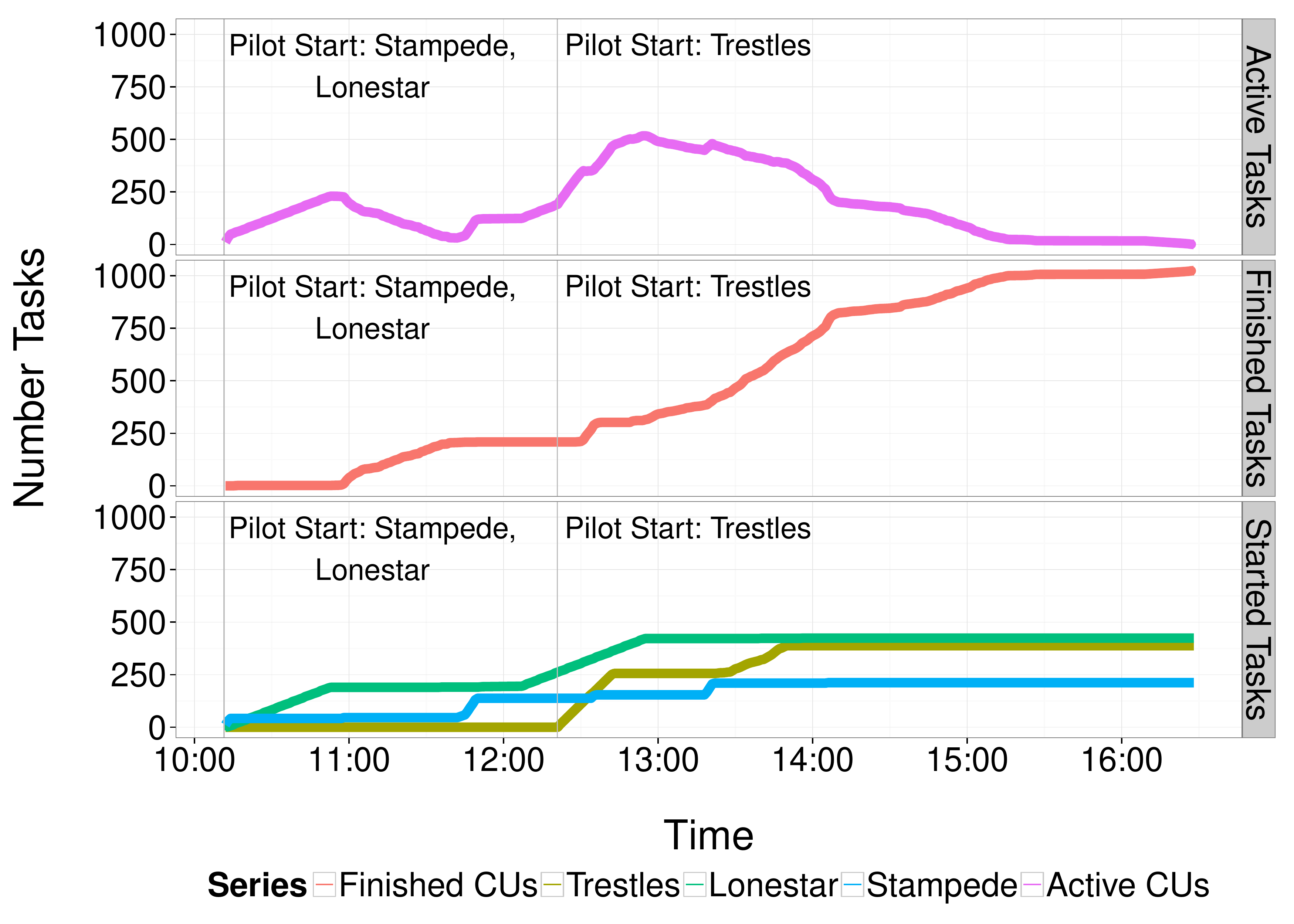}
	\caption{\textbf{Time series for a Single Run on 
	Lonestar/Stampede/Trestles:}  During this experiment 
	Stampede represented a significant bottleneck. Only 212 out of the 256 
	slots were claimed despite the fact that the \pilot was active from the 
	beginning. Trestles in contrast claimed only 36\,\cus less than Lonestar 
	despite the significantly larger \pilot queuing time.}
	\label{fig:performance_3way-timeseries}
\end{figure}

Finally, we explore the distribution of the workload across 3 machines in a
wide area network in scenario 4. Again, we utilize the \du replication
capability of \pilotdata. Several attempts of conducting the experiments
without replication failed. As seen in the
Figure~\ref{fig:performance_scaleout_bwa-1024cus-1GB-distribution}, the runtime
$T$ is about 6000\,sec longer than in the best case (scenario 3). Nevertheless,
it is still short than in the single resource scenario. 
Figure~\ref{fig:performance_3way-timeseries} shows the timeline of an 
example run. As indicated by the large error bars, the runtime of each CU 
fluctuates strongly: in general, CUs started later on a machine run longer. As 
seen, the number of active CUs is constrained by the non-availability of 
resources. After \pilot 3 becomes active the number of active CUs peaks.
Overall, we experienced a high fluctuation in the queue time on Trestles,
which also impacts both the distribution of \cus as well as the \cu runtime.
The more \cus that are allocated the Trestles, the slower the average runtime
of each \cu. As expected, with the degree of distribution the predictability of
the run decreases; minor differences (e.\,g.\ in the queue time) can
significantly alter the overall runtime. Thus, it is critical to deploy
application-level routines to react to dynamic changes in the resource
availability.

In summary, \pilot-Abstractions enable large-scale applications to use various
strategies for allocating distributed data and compute resources. The usage of
distributed resources enables applications to exploit additional resources in a
flexible manner, avoiding queuing times on a single resource. However, it must
be noted that in particular for long-running \cus, the first resource must not
be the best one. As seen in scenario 3, it can be beneficial to wait for the
faster machines even though there is a significant queueing time. Also,
particularly with larger ensembles, fault tolerance becomes a challenge. During
the runs we observed failures due to high loads (e.\,g.\ in scenario 1), wall
time limits and file transfer errors. In the future we will require
enhancements to BigJob in order to bolster fault tolerance (e.g. supporting a
reliable way to restart transfers). Also, more advanced strategies, such as
data replication and more fine granular partitioning, will improve the
distributability of large-scale applications.

\section{Discussion and Future Work}

An increasing number of scientific applications is data-driven, which is
associated with a new set of challenges for existing infrastructures and tools.
Science domains and applications greatly differ in the ways they generate,
store and use data. Managing applications with different characteristics on top
of heterogeneous resources at scale represents a serious challenge -- different
application workloads require different resource allocation and workload
placement strategies. Commonly, there are three primary compute-data placement 
paradigms:
(i) when the data is essentially localized, either from being ``poured'' into a
single storage backend or because the volumes of data allow for small-scale
localization; (ii) where the data is decomposed and distributed (with
multi-tier redundancy and caching) to an appropriate number of
computing/analytical engines as available, e.\,g., as employed by the EGI/OSG
for particle physics and the discovery of the Higgs, and (iii) a hybrid of the
above two paradigms, wherein data is decomposed and committed to several
infrastructures, which could then result in a combination of either of the
first two paradigms. Even though specific realizations and backends vary, we
have shown \pilotdata supports reasoning over different compute-data placement
paradigms and infrastructures. \pilot-Abstractions allow applications to map
system-level capabilities, to a unified, logical resource topology, which
enables the application to reason about trade-off and optimize placement
decisions accordingly based on the information provided by the affinity model,
such as resource localities, and dynamic information, such as resource and
bandwidth availabilities.

\pilotdata enhances the utility and usability of \pilotjobs by extending the
use of \pilot-Abstractions to data and thus, providing a missing critical
component. In conjunction with the fact that \pilotjobs provide a well-defined
abstraction for distributed resource management independent of
infrastructure-specific details, the combined \pilot-Abstraction of \pilotdata
and \pilotjob is a powerful approach to manage the compute and data challenges
in heterogeneous and highly dynamic distributed environments. With increasing
heterogeneity (e.\,g.\ when using multiple infrastructures as grids, clouds,
and HTC environments), the unpredictability and dynamisms increases as well.
Thus, it is critical to provide the right level of control that enables the
application to respond to this kind of dynamism, e.\,g.\ by acquiring
additional compute resources close to the replica of the dataset. Our genome
sequencing application for example successfully demonstrates how \pilotdata
provides the right primitives for expressing tasks and their data dependencies
and for exploring trade-offs such as data replication in distributed and
dynamic environments. We also successfully showed that the \pilotdata
efficiently supports other application patterns, e.\,g.\ dynamic
workflows~\cite{Pilot-Data-Workflow} or
MapReduce~\cite{Mantha:2012:PEF:2287016.2287020}.

The above functional and qualitative attributes exposed via \pilot-Abstractions
enable a simple method of managing complexity (inherent to working with data
across diverse infrastructures), thereby supporting the claim that \pilotdata
provides an unifying abstraction for distributed data and compute. The
advantages of \pilotdata, however, extend well beyond the conceptual: through a
series of experiments that cover a range of often-realized distributed
configurations and scenarios, we have seen how \pilotdata provide an
abstraction and a powerful tool for managing distributed data. We reiterate
that the application-scenarios investigated as well implementations are
production-grade and used on production DCI such as XSEDE, EGI and OSG, along
with their inherent complexity. We have shown that \pilotdata effectively can
distribute data and compute across these infrastructure utilizing system-level
features, such as \irods-based replication on OSG, helping the application to
optimize data/compute placement, e.\,g.\ by utilizing system-level support for
replication where available (e.\,g.\ on OSG) and deploy \pilotdata-level
replication in other cases (such as XSEDE).

In the future, we will explore the \pilot-Abstraction as a basis for building
higher-level capabilities and frameworks (e.\,g.\ for workload and workflow 
management) to provide further productivity gains. Multi-level scheduling
has been demonstrated to be an effective mean to address complex and diverse
application characteristics and associated requirements with respect to
resource management. Our discussion of affinities suggested that they are a
good abstraction for capturing relationships between computational tasks and
associated data and help to map these dependencies to \pilots. Our prototype
workload management service (the \computedataservice), which is based on a
simple affinity model, will become the basis for \pilotdata's enhanced
scheduling capabilities enabling dynamic execution decisions based on incoming
data or varying infrastructure conditions. For this purpose, we will explore
different, heterogeneous workloads, e.\,g.\ ensembles, data-intensive tasks,
workflows, etc. We will investigate the characteristics of these workloads to
derive important parameters such as runtime and data characteristics for
optimized scheduling. Further, we will investigate high-level abstractions for
re-occurring data/compute usage patterns, e.\,g.\ data partitioning, filtering
or merging of datasets etc. further improving developer productivity.

\section*{Acknowledgements}

This work is primarily funded by NSF CAREER Award (OCI-1253644), as
well as by NSF Cyber-enabled Discovery and Innovation Award
(CHE-1125332), NSF-ExTENCI (OCI-1007115), NSF EarthCube (SCIHM,
OCI-1235085) and Department of Energy Award (ASCR,
DE-FG02-12ER\-26115).  Pradeep Mantha (Berkeley) contributed to early
associated work on Pilot-Data based applications.  
We thank our fellow RADICAL members for
their support and input, in particular Andre Merzky, Matteo Turilli
and Melissa Romanus. We thank Tanya Levshina for help with iRODS
configurational issues on OSG. SJ acknowledges useful related
discussions with Jon Weissman (Minnesota) and Dan Katz (Chicago). SJ
acknowledges UK EPSRC via for supporting the e-Science Research
themes ``Distributed Programming Abstractions'' \& 3DPAS and earlier
support of the SAGA project. AL and SJ acknowledge NSF-OCI
1059635. This work has also been made possible thanks to computer
resources provided by TeraGrid TRAC award TG-MCB090174. We thank
Yaakoub El-Khamra of TACC for exceptional support on XSEDE
systems. EGI experiments were performed on resources provided by
SURFsara's Dutch e-Science Grid.


\end{document}